\documentclass[a4paper,USenglish,cleveref, autoref, thm-restate]{lipics-v2021}

\usepackage{multicol}
\usepackage{bbm}

\crefname{lemma}{Lemma}{Lemmas}
\crefname{claim}{Claim}{Claims}

\newcommand{\para}[1]{\subparagraph*{#1}}

\renewcommand{\L}{\mathcal{L}}
\newcommand{\B}{\mathcal{B}}
\newcommand{\R}{\mathcal{R}}
\newcommand{\N}{\mathbb{N}}

\renewcommand{\S}{\mathcal{S}}
\newcommand{\Pref}{\mathcal{P}}
\newcommand{\Suf}{\mathcal{S}}

\newcommand{\cT}{\mathcal{T}}
\newcommand{\cD}{\mathcal{D}}
\newcommand{\cX}{\mathcal{X}}

\newcommand{\preftype}{\mathsf{pref}}
\newcommand{\sufftype}{\mathsf{suff}}

\newcommand{\Valid}{\mathsf{Valid}}

\newcommand{\I}{\mathcal{I}}
\newcommand{\C}{\mathcal{C}}

\newcommand{\Dict}{\mathsf{Dict}}

\newcommand{\per}{\mathsf{per}}
\newcommand{\run}{\mathsf{run}}
\newcommand{\Run}{\mathsf{Run}}
\newcommand{\close}{\mathsf{close}}
\renewcommand{\root}{\mathsf{root}}
\newcommand{\Cover}{\mathsf{Cover}}

\newcommand{\Pairs}{\mathsf{Pairs}}

\newcommand{\pref}{\mathsf{pref}}
\newcommand{\suff}{\mathsf{suff}}

\renewcommand{\int}{\mathsf{int}}

\newcommand{\sub}{\mathsf{sub}}
\newcommand{\bor}{\mathsf{bor}}

\newcommand{\SufffExtend}{\mathsf{SuffExtend}}
\newcommand{\SubExtend}{\mathsf{SubExtend}}
\newcommand{\PrefExtend}{\mathsf{PrefExtend}}
\newcommand{\EndSymbol}{\mathsf{End}}

\newcommand{\occ}{\mathsf{occ}}
\newcommand{\LCP}{\mathsf{LCP}}
\newcommand{\CNF}{\mathsf{CNF}}
\newcommand{\Add}{\mathsf{Add}}
\newcommand{\pper}{\mathsf{pper}}
\newcommand{\paper}{\mathsf{paper}}
\newcommand{\sper}{\mathsf{sper}}
\newcommand{\saper}{\mathsf{saper}}
\newcommand{\subper}{\mathsf{subper}}
\newcommand{\subaper}{\mathsf{subaper}}

\newcommand{\IPM}{\mathsf{IPM}}

\newcommand{\RIGHT}{\mathsf{right}}
\newcommand{\LEFT}{\mathsf{left}}
\newcommand{\trim}{\mathsf{trim}}
\newcommand{\Output}{\mathsf{output}}

\newcommand{\ceil}[1]{\left\lceil{#1}\right\rceil}
\newcommand{\floor}[1]{\left\lfloor{#1}\right\rfloor}

\title{String 2-Covers with No Length Restrictions} 

\author{Itai Boneh}{Reichman University and University of Haifa, Israel}{itai.bone@biu.ac.il}{https://orcid.org/0009-0007-8895-4069}{supported by Israel Science Foundation grant 810/21.}

\author{Shay Golan}{Reichman University and University of Haifa, Israel}{golansh1@biu.ac.il}{https://orcid.org/0000-0001-8357-2802}{supported by Israel Science Foundation grant 810/21.}

\author{Arseny Shur}{Bar Ilan University, Israel}{shur@datalab.cs.biu.ac.il}{https://orcid.org/0000-0002-7812-3399}{supported by the ERC grant MPM under the EU's Horizon 2020 Research and Innovation Programme (grant no. 683064) and by the State of Israel through the Center for Absorption in Science of the Ministry of Aliyah and Immigration.}

\authorrunning{Boneh, Golan and Shur} 

\Copyright{Jane Open Access and Joan R. Public} 



\category{}
\relatedversion{} 

\hideLIPIcs  
\nolinenumbers
\begin{document}

\maketitle

\begin{abstract}
A $\lambda$-cover of a string $S$ is a set of strings $\{C_i\}_1^\lambda$ such that every index in $S$ is contained in an occurrence of at least one string $C_i$.
The existence of a $1$-cover defines a well-known class of quasi-periodic strings.
Quasi-periodicity can be decided in linear time, and all $1$-covers of a string can be reported in linear time plus the size of the output.
Since in general it is NP-complete to decide whether a string has a $\lambda$-cover,
the natural next step is the development of efficient algorithms for $2$-covers.
Radoszewski and Straszyński [ESA 2020] analysed the particular case where the strings in a $2$-cover must be of the same length. 
They provided an algorithm that reports all such $2$-covers of $S$ in time near-linear in $|S|$ and in the size of the output.

In this work, we consider $2$-covers in full generality. 
Since every length-$n$ string has $\Omega(n^2)$ trivial $2$-covers (every prefix and suffix of total length at least $n$ constitute such a $2$-cover), we state the reporting problem as follows:
given a string $S$ and a number $m$, report all $2$-covers $\{C_1,C_2\}$ of $S$ with length $|C_1|+|C_2|$ upper bounded by $m$.
We present an $\Tilde{O}(n + \Output)$ time algorithm solving this problem, with $\Output$ being the size of the output.
This algorithm admits a simpler modification
that finds a $2$-cover of minimum length.
We also provide an $\Tilde{O}(n)$ time construction of a $2$-cover oracle which, given two substrings $C_1,C_2$ of $S$, reports in poly-logarithmic time whether $\{C_1,C_2\}$ is a $2$-cover of $S$.
\end{abstract}
\clearpage
\setcounter{page}{1} 

\section{Introduction}\label{sec:intro}

For a string $S$, the substring $C$ of $S$ is a \textit{cover} of $S$ if every index of $S$ is covered by an occurrence of $C$.
Since the introduction of covers by Apostolico and Ehrenfeucht~\cite{AE93}, many algorithms have been developed for finding covers or variations of covers of a given string.
~\cite{AE93} presented an $O(n\log^2 n)$ time algorithm for finding all covers of an input string of length $n$.
It was shown by Moore and Smyth~\cite{MS94,MS95} that all covers of a string can be reported in $O(n)$ time.
Smyth~\cite{S02} further extended this result by showing that the covers of all prefixes of $S$ can be computed in $O(n)$ time.
Further works on covers and variants of cover include \cite{ARS16,KPRRW15,BIP95,ALLLP19,ALLP19,GRS19,BKLPR20,PT19}.

A natural generalization of a cover is a $\lambda$-cover.
A set of strings $\{C_1,C_2, \ldots, C_\lambda\}$ is a $\lambda$-cover of $S$ if every index in $S$ is covered by an occurrence of $C_i$ for some $i \in [\lambda]$.
The notion of $\lambda$-covers was introduced by Guo, Zhang and Iliopoulos~\cite{GZI07,GZI08}, who proposed an $O(n^2)$ time algorithm for computing all $\lambda$-covers of a string $S$ over a constant size alphabet for a given $\lambda$.
The running time analysis of their algorithm was later found to be faulty by Czajka and Radoszewski~\cite{CR21}, who showed that it has an exponential worst-case running time.
Cole et al.~\cite{CIMS05} justified the lack of a polynomial algorithm for computing all $\lambda$-covers by proving that the problem is NP-complete.

In this work, we focus on $2$-covers.
Radoszewski and Straszy{\'n}ski~\cite{RS20} have considered a special case of ``balanced'' $2$-covers, where the two strings composing the $2$-cover have equal length.
They proposed an $\Tilde{O}(n + \Output)$-time algorithm reporting all balanced $2$-covers of a given length-$n$ string $S$.
They also provide two versions of this algorithm, one of which finds a balanced $2$-cover of each possible length and the other determines the shortest balanced $2$-cover of $S$; both versions work in $\Tilde{O}(n)$ time.

Designing efficient algorithms for the same problems on $2$-covers in the general case is posed in~\cite{RS20} as an open problem.

For a $2$-cover $\{C_1,C_2\}$, its \emph{length} is $|C_1|+|C_2|$.
Following the open problem of Radoszewski and Straszy{\'n}ski, we specify the following problems for $2$-covers in the general case:
\begin{itemize}
    \item \textsf{All\_2-covers}$(S,m)$: for a string $S$, report all $2$-covers of length at most $m$;
    \item \textsf{Shortest\_2-cover}$(S)$: for a string $S$, find a $2$-cover of minimum length;
    \item \textsf{2-cover\_Oracle}$(S)$: for a string $S$, build a data structure that answers the queries of the form ``do given two substrings of $S$ constitute a $2$-cover of $S$?''
\end{itemize}

Note that every length-$n$ string $S$ trivially has $\Omega(n^2)$ $2$-covers.
E. g., if $C_1$ is a prefix of $S$, $C_2$ is a suffix of $S$, and $|C_1|+|C_2|\ge n$, then $\{C_1,C_2\}$ is a $2$-cover of $S$.
The length restriction made in the above formulation of the \textsf{All\_2-covers} problem allows one to consider instances with smaller outputs, thus returning the running times of type $\Tilde{O}(n + \Output)$ into the game.

\para{Our contribution.}
In this work, we solve the three above problems in near linear time.
\begin{theorem}\label{thm:reporting}
    There exists an algorithm that solves \emph{\textsf{All\_2-covers}$(S,m)$} in $O(n\log^{5}n+\Output\cdot \log^{3}n)$ time.
\end{theorem}
\begin{theorem}\label{thm:shortest}
    There exists an algorithm that solves \emph{\textsf{Shortest\_2-cover}$(S)$} in $O(n\log^{4}n)$ time. 
\end{theorem}
\begin{theorem}\label{thm:oracle}
    There exists an algorithm that solves \emph{\textsf{2-cover\_Oracle}$(S)$} in $O(n \log^5n)$ preprocessing time and $O(\log^3 n)$ query time.
\end{theorem}

\para{Techniques and Ideas.}
The main idea of our algorithms is the formulation of the property ``an index $i$ in $S$ is covered by an occurrence of a substring $U$ of $S$'' in terms of point location.
At a high level, each index $i$ is assigned a compactly representable area $\mathcal{A}_i$ in the plane, and every substring $U$ that is \emph{not highly periodic} corresponds to a point $p_U$ in the plane such that $p_U \in \mathcal{A}_i$ if and only if the index $i$ is covered by some occurrence of $U$.
The same idea, implemented in three dimensions instead of two, covers the case of highly periodic substrings.

Given such a geometric representation, our algorithms make use of multi-dimensional range-reporting and range-stabbing data structures to retrieve and organize the areas associated with each index in $S$. 
This organization facilitates the computation of a \textit{core} set of the $2$-covers, which consists of pairs of strings that are not highly periodic.
This set provides a solution to the \textsf{Shortest\_2-cover} problem. 
Besides that, we create the oracle solving the \textsf{2-cover\_Oracle} problem and utilize the core set to finalize our solution to the \textsf{All\_2-covers} problem with a small number of queries to this oracle.

\para{Organization.}
In \cref{sec:prelim} we present notation, auxiliary lemmas, and pre-existing data structures that are used in our algorithms.
In \cref{sec:ranges} we formalize and prove the connection between covering an index in a string by an occurrence of a substring and multidimensional point location.
In \cref{sec:oracle} we build upon the insights presented in \cref{sec:ranges} to design the $2$-cover oracle and prove \cref{thm:oracle}.
Finally, in \cref{sec:report} we present the reporting algorithms proving \cref{thm:shortest} and \cref{thm:reporting} (the latter one executes the oracle).
All details omitted due to space constraints are put into Appendix.

\section{Preliminaries}\label{sec:prelim}
Here we present definitions, notation, and auxiliary lemmas.
For completeness, proofs of the statements appearing in this section are given in \cref{app:pre}.

We assume in this paper that $0\in\N$.
We denote $[x..y]=\{i\in\N\mid x\le i\le y\}$ for any \emph{real} numbers $x,y$, possibly negative.
We also denote $[x]=[1..x]$.
The notation $\Output$ stands for the size of output of a reporting algorithm.

All strings in the paper are over an alphabet $\Sigma=\{1,2,\ldots O(n^c)\}$ for some constant $c$.
The letters of a string $S$ are indexed from 1 to $|S|$.
If $X=S[i..j]$, $X$ is called a \emph{substring} of $S$ (a \emph{prefix} of $S$ if $i=1$, a \emph{suffix} of $S$ if $j=|S|$, and an empty string if $i>j$).
We also say that $S[i..j]$ specifies an \emph{occurrence of $X$ at position $i$}.
If $X$ is a substring of $S$, then $S$ is a \emph{superstring} of $X$. 
A string $X$ that occurs both as a prefix and as a suffix of $S$ is a \emph{border} of $S$.
A string $S$ has \emph{period} $\rho$ if $S[i]=S[i+\rho]$ for all $i\in[|S|-\rho]$.
Clearly, $S$ has period $\rho$ if and only if $S[1..|S|-\rho]$ is a border of $S$.
The minimal period of $S$ is denoted by $\per(S)$.
Let $\per(S)=\rho$.
We say that $S$ is \emph{aperiodic} if $|S|<2\rho$, \emph{($\rho$-)periodic} if $|S|\ge2\rho$, and \emph{highly ($\rho$-)periodic} if $|S|\ge3\rho$.
We say that $S$ is short ($\rho$)-periodic if $|S| \in [2\rho .. 3\rho -1]$. 
Note that if a string $X$ occurs in $S$ at positions $i$ and $j$, then $|j-i|\ge \per(X)$.

The following two lemmas specify some useful structure of periodic prefixes and borders.

\begin{lemma}\label{lem:fewperiods}
The prefixes of a length-$n$ string have, in total, $O(\log n)$ different periods.
\end{lemma}

\begin{lemma}\label{lem:fewborders}
    Every string of length $n$ has $O(\log n)$ aperiodic borders and $O(\log n)$ short periodic borders.
\end{lemma}

We use well known notion of the longest common prefix ($\LCP$).

\begin{definition}
    For two strings $S$ and $T$, 
    $\LCP(S,T)=\max\{\ell\mid S[1..\ell]=T[1..\ell]\}$ is the length of their longest common prefix and $\LCP^R(S,T)=\max\{\ell\mid S[|S|-\ell+1 ..|S|]=T[|T|-\ell+1..|T|]\}$ is the length of their longest common suffix.
\end{definition}

\para{Covers.}Given a string $S$, we say that a
substring $X$ \emph{covers} an index $i$ if for some indices $j_1,j_2\in[|S|]$ we have $X=S[j_1..j_2]$ and $i\in[j_1..j_2]$. 
We also say that the occurrence of $X$ at $j_1$ covers $i$.
If $X$ covers every $i\in[|S|]$, we call $X$ a \emph{$1$-cover} of $S$.
A pair of substrings $(X,Y)$ is said to cover $i$ if $X$ or $Y$ covers $i$. 
If a pair $(X,Y)$ covers every $i\in[|S|]$, we call $(X,Y)$ a \emph{$2$-cover} of $S$.
(It will be convenient to consider 2-covers as ordered pairs, though the notion of 2-cover is symmetric with respect to $X$ and $Y$.)
We say that $(X,Y)$ is highly periodic if either $X$ or $Y$ is highly periodic, otherwise $(X,Y)$ is non-highly periodic. The following lemma considers a periodic string that covers index $i$. 

\begin{lemma}\label{lem:removePeriod}
    Let $S$ be a string, and let $X$ be a $\rho$-periodic substring.
    If the string $X$ covers an index $i$, then the string $X[1..|X|-\rho]]$ (= $X[\rho+1 .. |X|]$) also covers $i$.
\end{lemma}

\para{Runs.}
A $\rho$-periodic substring of $S$ is a \emph{run} if it is not contained in a longer $\rho$-periodic substring.
We use the following lemmas regarding runs.

\begin{lemma}\label{lem:runscover}
Let $S$ be a string, $\rho\in[|S|]$.
Every index $i$ in $S$ is covered by at most two $\rho$-periodic runs and by $O(\log n)$ highly-periodic runs.
\end{lemma}

\begin{lemma}\label{lem:uniquerun}
    Let $S$ be a string. For every $\rho$-periodic substring $S[i..j]$, there is a unique $\rho$-periodic run containing $S[i..j]$.
\end{lemma}

\begin{lemma}\label{lem:runextaper}
    If there is an integer $\rho$ such that $S[x..y]$ is $\rho$-periodic and $S[x..y+1]$ is not $\rho$-periodic, then $S[x..y+1]$ is aperiodic. 
\end{lemma}

\begin{lemma}[{\cite[{Theorem 9}]{BIINTT17}}]\label{fact:runtheorem}
The number of runs in any string $S$ is smaller than $|S|$.
\end{lemma}

\subsection{Range Data Structures}

Our algorithms use data structures for \emph{orthogonal range queries}.
Such a data structure is associated with a positive integer dimension $d$ and deals with $d$-dimensional points and $d$-dimensional ranges.
A $d$-dimensional point is a $d$-tuple $p = (x_1,x_2, \ldots , x_d)$ and a $d$-dimensional range is the cartesian product $R = [a_1 .. b_1] \times [a_2 .. b_2] \times \ldots \times[a_d .. b_d]$ of $d$ ranges.
We call a $2$-dimensional range a \emph{rectangle} and a $3$-dimensional range a \emph{cuboid}.
We say that a point $p$ is contained in the range $R$ (denoted by $p \in R$) if
$x_i \in [a_i .. b_i]$ for every $i\in [d]$. 

We make use of the following range data structures.

\begin{lemma}[Range Query Data Structure~\cite{W85,Chazelle88}]\label{lem:rangeds}
For any integer $d$, a set $P$ of $n$ points in $\mathbb{R}^d$ can be preprocessed in $O(n\log^{d-1}n)$ time to support the following queries. 
\begin{itemize}
    \item \textbf{Reporting:} Given a $d$-dimensional range $R$, output all points in the set $P\cap R$.
    \item \textbf{Emptiness:} Given a $d$-dimensional range $R$, report if $P\cap R = \emptyset$ or not.
\end{itemize}  
The query time is $O(\log^{d-1} n)$ for Emptiness and $O(\log^{d-1} n + \Output)$ for Reporting.
\end{lemma}

\begin{lemma}[{Range stabbing queries \cite[{Theorems 5 and 7}]{Chazelle88}}]\label{lem:stab}
    For any integer $d$, a set of $d$-dimensional ranges $R_1,R_2,\ldots R_n$ can be prepossessed in $O(n\log^{d-1}n)$ time to support the following queries.
    \begin{itemize}
        \item \textbf{Stabbing:} Given a $d$-dimensional point $p$, report all ranges $R_i$ such that $p\in R_i$.
        \item     \textbf{Existence:} Given a $d$-dimensional point $p$, report if \textbf{no} ranges $R_i$ satisfy $p\in R_i$.
    \end{itemize}
    The query time is $O(\log^{d-1}n)$ for Existence and $O(\log^{d-1}n + \Output)$ for Stabbing. 
\end{lemma}

\subsection{Stringology Algorithms and Data Structures}
Throughout the paper, we make use of the following string algorithms and data structures.

\begin{lemma}[Pattern Matching~\cite{KMP77}]\label{lem:PM}
There exists an algorithm that, given a string $T$ of length $n$ and a string $P$ of length $m\le n$, reports in $O(n)$ time all the occurrences of $P$ in $T$.
\end{lemma}

\begin{lemma}[$\LCP_S$ Data Structure~\cite{LV88,GG88}]\label{lem:LCP}
    There exists a data structure $\LCP_S$ that preprocesses an arbitrary string $S\in\Sigma^*$ 
    of length $n$ in $O(n)$ time and supports constant-time queries $\LCP_S(i,j)=\LCP(S[i..n],S[j..n])$ and $\LCP^R_S(i,j)=\LCP^R(S[1..i],S[1..j])$.
\end{lemma}

When $S$ is clear from context, we simply write $\LCP(i,j)$ and $\LCP^R(i,j)$.

\begin{lemma}[Internal Pattern Matching ($\IPM$)~\cite{KRRW13}]\label{lem:IPM}
    There exists a data structure $\IPM_S$ that preprocesses an arbitrary string $S\in\Sigma^*$ of length $n$ in $O(n)$ time and supports the following constant-time queries.
\begin{itemize}
    \item \textbf{Periodic:} given a substring $X$, return $\per(X)$ if $X$ is periodic, and ``aperiodic'' otherwise.

    \item \textbf{Internal Matching:} Given two substrings $X$ and $Y$ such that $|Y| = O(|X|)$, return all occurrences of $X$ in $Y$ represented as $O(1)$ arithmetic progressions.
\end{itemize}    
\end{lemma}

\begin{lemma}[Finding all Substrings, see \cite{LT22}]\label{lem:allsubstrings}
    There is an algorithm that reports all  distinct substrings of a string $S[1..n]$ in time $O(n + \Output)$.
\end{lemma}

\begin{lemma}[{Finding all Runs~\cite[{Theorem 1.4}]{EGG23}}]\label{lem:runs}
    There is algorithm that computes all runs of a string $S\in\Sigma^*$ of length $n$ in $O(n)$ time.
\end{lemma}

\section{Range Characterization of Covering an Index}\label{sec:ranges}

In this section we translate the property ``an index is covered by an occurrence of a given substring'' to the language of $d$-dimensional points and ranges.
Then this property can be checked with the queries described in \cref{lem:rangeds,lem:stab}.
We distinguish between the $2$-dimensional case of not highly periodic substrings (\cref{lem:rangecoveraperRuntime}) and 3-dimensional case of highly periodic substrings (\cref{lem:rangecoverperRuntime}).
Given a point $p$ and a set $\R$ of ranges (both in $d$ dimensions), we slightly abuse the notation, writing $p\in\R$ instead of $p\in\bigcup_{R\in\R} R$.

To present the algorithms that prove these lemmas, we first describe an $O(n\log^2n)$ time preprocessing phase.
Throughout the rest of the paper, we assume that this preprocessing has already been executed.

\para{Preprocessing.}
The algorithm computes $\LCP_S$ data structure of \cref{lem:LCP} and $\IPM_S$ data structure of \cref{lem:IPM}.
In addition, the algorithm computes all runs of $S$ using \cref{lem:runs}.
For every $\rho\in[n]$, the algorithm stores all $\rho$-periodic runs of $S$ in a $3$-dimensional range reporting data structure $D^\rho_\run$ of \cref{lem:rangeds} as follows.
For every $\rho$-periodic run $S[\ell..r]$, the data structure $D^\rho_\run$ contains the point $p=(\ell,r,r-\ell+1)$.
By \cref{fact:runtheorem}, the total number of runs stored in the structures $D^\rho_\run$ over all $\rho \in [n]$ is at most $n$.
It follows from \cref{lem:LCP,lem:IPM,lem:runs,lem:rangeds} that the preprocessing time is $O(n\log^2n)$.

\subsection{The Not Highly Periodic Case}
In this section we prove the following lemma.

\begin{lemma}
\label{lem:rangecoveraperRuntime}
    Let $f,i\in[n]$ be two indices and let $k\in\N$.
    There exists a set $\R$ of $O(1)$ rectangles such that for any $\ell,r\in\N$ with $\ell+r+1\in [1.5^k..1.5^{k+1}]$ the string $\sub=S[f-\ell..f+r]$ satisfies the following conditions:
    \begin{enumerate}
        \item If $(\ell,r) \in\R$, then $\sub$ covers $i$ and $\per(\sub) \ge \frac{1.5^k}{4}$.
        \item If $\sub$ covers $i$ and is not highly periodic, then $(\ell,r) \in\R$.
    \end{enumerate}
    Moreover, $\R$ can be computed in $O(\log^2 n)$ time.
\end{lemma}
For $f\in [n]$ and $k\in\N$, let $\sub_\LEFT = S[f-\floor{\frac{1.5^k}{2}}..f]$, and $\sub_\RIGHT =S[f..f+\floor{\frac{1.5^k}{2}}]$. If an endpoint of a substring is outside $S$, the substring is undefined.

\begin{observation}\label{obs:covconrorl}
Let $f\in [n]$ be an index and $k\in\N$.
For every $\sub=S[f-\ell..f+r]$ such that $\ell,r \in \N$ and $ |\sub|\in[1.5^k..1.5^{k+1}]$, $\sub$ is a superstring of either $\sub_{\LEFT}$ or $\sub_{\RIGHT}$.
\end{observation}

\cref{obs:covconrorl} allows us to prove \cref{lem:rangecoveraperRuntime} as follows.
First we find a set $\R_1$ that satisfies the conditions of the lemma for all pairs $(\ell,r)$ such that $\sub = S[f-\ell .. f+r]$ is a superstring of $\sub_\RIGHT$.
Similarly, we find a set $\R_2$ for the case where $\sub$ is a superstring of $\sub_\LEFT$.
Then $\R_1\cup \R_2$ is the set required by the lemma.

In the rest of the section we show how to find the set $\R_1$. 
(The argument for the set $\R_2$ is similar, so we omit it.) 
Let us fix $f,\ell$, and $r \ge \floor{\frac{1.5^k}{2}} = |\sub_{\RIGHT}|-1$.
Let $i_{\RIGHT}$ denote the starting index of an occurrence of $\sub_\RIGHT$.
We make the following claim.
\begin{claim}\label{clm:bsvalidrectangle}
    There exists a rectangle $R$ such that for any $\ell,r\in\N$ with $\ell+r+1\in [1.5^k..1.5^{k+1}]$ and $r\ge \floor{\frac{1.5^k}{2}}$ the substring $\sub = S[f-\ell .. f+r]$ covers the index $i$ \textbf{with the occurrence at position $i_\RIGHT-\ell$} if and only if $(\ell,r) \in R$.
    Moreover, $R$ can be computed in $O(1)$ time.
\end{claim}
\begin{claimproof}
    Let $e_r = \LCP(f,i_{\RIGHT})$, $e_\ell = \LCP^R(f,i_{\RIGHT})$.
    Using $\LCP_S$, we compute in $O(1)$ time the rectangle $R = [i_{\RIGHT} - i\:..\:e_\ell-1]\times[i - i_{\RIGHT}\:..\:e_r-1]$ and check the required conditions.
    
    First assume $(\ell,r) \in R$.
    Since $\ell \le e_\ell-1$ and $r \le e_r-1$, one has $\sub = S[f-\ell..f+r] = S[i_\RIGHT-\ell..i_\RIGHT+r]$, so $\sub$ occurs at $i_{\RIGHT} - \ell$.
    Since $\ell \ge i_\RIGHT-i$ and $r \ge i-i_\RIGHT$, we have $i \in [i_\RIGHT - \ell .. i_\RIGHT + r]$, so this occurrence covers $i$.

    Now assume that $\sub$ covers $i$ with the occurrence at $i_\RIGHT - \ell$.
    Since $\sub$ occurs at $i_\RIGHT - \ell$, one has $\ell \le e_\ell-1$ and $r \le e_r-1$.
    Since this occurrence covers $i$, one also has $i \in  [i_\RIGHT-\ell .. i_\RIGHT - \ell + |\sub| - 1]=[i_\RIGHT - \ell .. i_\RIGHT + r]$.
    Then $\ell \ge i_\RIGHT-i$ and $r \ge i-i_\RIGHT$, which finally proves $(\ell,r) \in R$.
\end{claimproof}

If $\sub$ covers index $i$, its substring $\sub_\RIGHT$ must occur close to $i$.
Since $|\sub|\le 1.5^{k+1}$, if an occurrence of $\sub$ at $i_\RIGHT-\ell$ covers $i$, then the position $i_\RIGHT$, at which  $\sub_\RIGHT$ occurs, is inside the range  $[i-1.5^{k+1}..i+ 1.5^{k+1}]$.
Let $\occ_{\RIGHT}$ be the set of all such indices $i_\RIGHT$ from this range.
We distinguish between two cases, regarding the period of $\sub_\RIGHT$.

\para{Case 1: $\per(\sub_{\RIGHT})\ge \frac{1.5^{k}}{4}$.}
The following claim is easy.
\begin{claim}\label{clm:bsfewrelevantaper}
    If $\per(\sub_{\RIGHT}) \ge \frac{1.5^k}{4}$, then
    $|\occ_{\RIGHT}|=O(1)$.
\end{claim}
\begin{claimproof}
The distance between two consecutive occurrences of $\sub_\RIGHT$ is at least $ \frac{1.5^k}{4}$.
Since a range of length $2\cdot1.5^{k+1}$ contains $O(1)$ disjoint ranges of length $ \frac{1.5^k}{4}$, the claim follows.
\end{claimproof}

Now we build the set $\R_1$.
We compute $\occ_{\RIGHT}$ in $O(1)$ time using the $\IPM_S$ data structure. 
For every $j \in \occ_{\RIGHT}$, we take the rectangle $R_j$ from \cref{clm:bsvalidrectangle}.
Let $\R_1=\{R_j\mid j\in \occ_{\RIGHT}\}$.
By \cref{clm:bsfewrelevantaper}, $|\R_1|=O(1)$.
Consider a pair $(\ell,r)$ such that $\ell+r+1\in [1.5^k..1.5^{k+1}]$ and $r\ge \floor{\frac{1.5^k}{2}}$, and let $\sub = S[f-\ell .. f+r]$.
If $(\ell,r) \in\R_1$, then $(\ell,r)\in R_j$ for some $j\in\occ_\RIGHT$.
Hence by \cref{clm:bsvalidrectangle} $\sub$ covers $i$ with an occurrence at $j-\ell$ and $\per(\sub)\ge\per(\sub_\RIGHT)\ge \frac{1.5^k}{4}$, as required.
Conversely, if $\sub$ covers $i$ with an occurrence at $j$, then $j'=j+\ell$ belongs to $\occ_\RIGHT$.
Then by \cref{clm:bsvalidrectangle} $(\ell,r)\in R_{j'}$ and thus $(\ell,r)\in\R_1$.
This concludes the proof of \cref{lem:rangecoveraperRuntime} in the case $\per(\sub_{\RIGHT}) \ge \frac{1.5^k}{4}$.

\para{Case 2: $\rho=\per(\sub_{\RIGHT})< \frac{1.5^{k}}{4}$.}
Let $\run_f = S[f - \ell^f_\run.. f+ r^f_\run]$ be the $\rho$-periodic run containing $S[f..f+|\sub_{\RIGHT}|-1]$ (such a run exists by \cref{lem:uniquerun}).
Consider a pair $(\ell,r)$ such that 
$\ell+r+1\in [1.5^k..1.5^{k+1}]$ and $r\ge \floor{\frac{1.5^k}{2}}$, and let the string $\sub = S[f-\ell .. f+r]$ be not highly periodic.
Then $\per(\sub)>\frac{1.5^k}{3}>\rho$.
Hence $\sub$ is not a substring of $\run_f$, which means that either $\ell> \ell^f_\run$ or $r > r^f_\run$.
Below we assume $r > r^f_\run$; the other case is symmetric. 
We first observe that this inequality guarantees that $\per(\sub)$ is big enough.

\begin{claim}\label{clm:extensionnonperiodic}
    For every $\sub = S[f-\ell ..f+r]$ with $r > r^f_\run$ one has $\per(\sub) \ge \frac{1.5^k}{4}$.
\end{claim}\begin{claimproof}
    The substring $S[f..f+r^f_\run]$ is $\rho$-periodic and  $u=S[f..f+r^f_\run + 1]$ is not (otherwise, $\run_f$ is not a run).
    By \cref{lem:runextaper}, $u$ is aperiodic.
    Then $\per(u)> \frac{|u|}{2}\ge \frac{1.5^k}{4}$.
    It remains to note that $u$ is a substring of $\sub$.
\end{claimproof}

Let $\sub_\RIGHT^\to=S[f..f+r_\run^f]$.
Note that $\sub_\RIGHT^\to$ is a $\rho$-periodic suffix of the $\rho$-periodic run $\run^f$ and $\sub$ contains $\sub_\RIGHT^\to$ followed by a letter that breaks the period $\rho$.
This means that if $\sub$ covers $i$, then $S$ contains, close to $i$, a $\rho$-periodic run with the suffix $\sub_\RIGHT^\to$.
Let us say that a $\rho$-periodic run $S[a_\run\: ..\: b_\run]$ is \textit{close to $i$} if $a_\run \le i + 1.5^{k+1}$ and $b_\run \ge i-1.5^{k+1}$. 
Clearly, if $\sub$ covers $i$, it contains the suffix $\sub_\RIGHT^\to$ of a run close to $i$. 

Let $\Run_{\close}$ be the set of $\rho$-periodic runs close to $i$ with length at least $|\sub_\RIGHT^\to|$.
\begin{claim}\label{clm:fewcloseruns}
    $|\Run_\close|=O(1)$.
    Moreover, $\Run_\close$ can be computed in $O(\log^2n)$ time.
\end{claim}
\begin{claimproof}
    Assume that $\Run_\close$ is ordered by the positions of runs.
    Each of these runs has length at least $\frac{1.5^k}{2}$ and any two $\rho$-periodic runs overlap by less than $\rho< \frac{1.5^k}{4}$ positions.
    Then the positions of any two consecutive runs from $\Run_\close$ differ by more than $\frac{1.5^k}{4}$ and any two non-consecutive runs are disjoint.
    Since the first run ends no later than the position $i-1.5^{k+1}$ by definition of being close to $i$, the third and all subsequent runs start after this position.
    Again by definition, all runs start before the position $i+1.5^{k+1}$.
    The range $[i-1.5^{k+1}..i+1.5^{k+1}]$ contains $O(1)$ positions such that any two of them differ by more than $\frac{1.5^k}{4}$.
    Hence we get $|\Run_\close|=O(1)$.

    Querying $D^\rho_\run$ with the range $[-\infty .. i+1.5^{k+1}] \times [i-1.5^{k+1}..\infty]\times [|\sub_\RIGHT^\to| .. \infty]$ we get 
    all $\rho$-periodic runs that are close to $i$ (due to the first two coordinates) and have length at least $|\sub_\RIGHT^\to|$ (due to the last coordinate); i.e., what we get is $\Run_\close$.
    The query time is $O(\log^2n)$ by \cref{lem:rangeds}.
\end{claimproof}

Now we construct the set $\R_1$.
We query $D^{\rho}_\run$ with $[-\infty..f]\times[f+\floor{\frac{1.5^k}{2}}..\infty]\times[-\infty..\infty]$ to get the unique $\rho$-periodic run $\run_f = [f-\ell^f_\run .. f+r^f_\run]$ containing the substring $\sub_\RIGHT=S[f.. f+ \floor{\frac{1.5^k}{2}}]$.
Then we compute the $O(1)$-size set $\Run_\close$ (\cref{clm:fewcloseruns}).
For every $\run \in \Run_\close$ we check, with an $\LCP$ query, whether $\sub_\RIGHT^\to$ is a suffix of $\run$.
If yes, we compute the position $i_\RIGHT$ of this suffix from the parameters of the run. 
Since $i_\RIGHT$ is the position of an occurrence of $\sub_\RIGHT$, we apply \cref{clm:bsvalidrectangle} to obtain a rectangle $R=[\ell_1,\ell_2]\times[r_1,r_2]$.
Since in our argument we assume $r>r_\run^f$, we replace $r_1$ with $\max\{r_1, r_\run^f+1\}$. 
If the range for $r$ remains nonempty, we denoted the obtained rectangle by $R_\run$.

Let $\R_{\RIGHT} = \{R_\run \mid \run \in \Run_\close  \}$.
In a symmetric way, we consider the case $\ell> \ell_\run^f$ and build the set $\R_{\LEFT}$.
Finally we set $\R_1=\R_\RIGHT \cup \R_\LEFT$.
The time complexity is dominated by $O(1)$ queries to $D^{\rho}_\run$, which take $O(\log^2 n)$ by \cref{lem:rangeds}.

Now consider a pair $(\ell,r)$ such that $\ell+r+1\in [1.5^k..1.5^{k+1}]$ and $r\ge \floor{\frac{1.5^k}{2}}$, and let $\sub = S[f-\ell .. f+r]$.
If $(\ell,r) \in\R_1$, then $(\ell,r)$ belongs to some rectangle from $\R_\RIGHT$ or $\R_\LEFT$; these cases are symmetric, so let this rectangle be $R_\run\in \R_\RIGHT$, where $\run=[a_\run.. b_\run]$.
Hence by \cref{clm:bsvalidrectangle} $\sub$ covers $i$ with an occurrence at $b_\run-r_\run^f-\ell$.
We also have $\per(\sub)\ge \frac{1.5^k}{4}$ by \cref{clm:extensionnonperiodic}. 
Conversely, if $\sub$ is not highly periodic, then either $r>r_\run^f$ or $\ell>\ell_\run^f$. 
Without loss of generality, let $r>r_\run^f$. 
Now if $\sub$ covers $i$ with an occurrence at $j$, then there is an occurrence of $\sub_\RIGHT^\to$ at $j'=j+\ell$ that is a suffix of a $\rho$-periodic run $\run$.
Then by \cref{clm:bsvalidrectangle} we have $(\ell,r)\in R_\run$ and thus $(\ell,r)\in\R_1$.
Thus, we finished the proof of \cref{lem:rangecoveraperRuntime}.

\subsection{The Highly Periodic Case}

In this section we prove \cref{lem:rangecoverperRuntime}, which is the analog of \cref{lem:rangecoveraperRuntime} for periodic strings.

We begin with more notation.
Let $u$ be a $\rho$-periodic string having a substring $v$ of length $\rho$.
Then there exist unique integers $d_1,d_2\in[0..\rho-1]$ and $q\in \N$ such that $u=v[\rho{-}d_1{+}1..\rho]v^qv[1..d_2]$.
We abbreviate this representation as $u=v^{[d_1;q;d_2]}$.
\begin{observation} \label{obs:dldr}
    Let $u=v^{[d_1;q;d_2]}$.
    The numbers $d_1, d_2$, and $q$  can be computed in $O(1)$ time given $|u|,|v|$, and the position of any occurrence of $v$ in $u$.
\end{observation}
For a $\rho$-periodic substring $\sub=S[f-\ell..f+r]$, we define its \emph{root} by
\[
    \root=\begin{cases}
    S[f..f+\rho-1], & \text{if }r\ge \rho-1,\\
    S[f-\rho..f-1], &\text{otherwise}.
    \end{cases}
\] 
Thus, $\sub=\root^{[d_\ell;q_{\ell,r};d_r]}$ for some unique integers $d_\ell,d_r\in[0..\rho-1]$ and $q_{\ell,r}>0$.

\begin{lemma}\label{lem:rangecoverperRuntime}
    Let $f,i\in[n]$ be two indices and let $\rho\in [n]$.
    There exists a set $\C$ of $O(1)$ cuboids such that every highly $\rho$-periodic substring of the form $\sub=S[f-\ell..f+r]$ with $\ell,r\in\N$ satisfies the following: $\sub$ covers index $i$ if and only if $(d_\ell,d_r,q_{\ell,r})\in\C$.
    Moreover, $\C$ can be computed in $O(\log^2 n)$ time.
\end{lemma}

Let $\sub = S[f-\ell ..f+r]$ be highly $\rho$-periodic.
By \cref{lem:uniquerun}, each occurrence of $\sub$ is contained in a unique $\rho$-periodic run.
By \cref{lem:runscover}, there are at most two such runs containing $i$ (say, $\run_1$ and $\run_2$).
Hence if $\sub$ covers $i$, it does so with an occurrence contained either in $\run_1$ or in $\run_2$.
Then \cref{lem:rangecoverperRuntime} follows from \cref{lem:runinrun} below: we query $D^\rho_\run$ to get $\run_1$ and $\run_2$ (in $O(\log^2 n)$ time by \cref{lem:rangeds}), take the sets $\C_1$ and $\C_2$ given by \cref{lem:runinrun} for $\run_1$ and $\run_2$ respectively, and let $\C=\C_1\cup \C_2$.

\begin{lemma}\label{lem:runinrun}
    Let $\run = S[a_\run .. b_\run]$ be a $\rho$-periodic run containing $i$.
    There is a set $\C$ of $O(1)$ cuboids such that $\sub$ covers $i$ \textbf{with an occurrence contained in} $\run$ if and only if $(d_\ell,d_r,q_{\ell,r}) \in \C$.
    Moreover, the set $\C$ can be computed in $O(1)$ time.
\end{lemma}

In the rest of the section we describe the algorithm computing the set $\C$ of \cref{lem:runinrun}.

First the algorithm checks the length of $\run$.
Since $\sub$ is highly $\rho$-periodic, $\sub\ge 3\rho$.
If $|\run|<3\rho$, then $\sub$ has no occurrences in $\run$.
Hence in this case $\C=\varnothing$.
Next, the algorithm verifies if $\root$ is a substring of $\run$.
Due to $\rho$-periodicity of $\run$, it is sufficient to check for an occurrence of $\root$ the prefix of $\run$ having length $2\rho = 2 |\root|$. 
By \cref{lem:IPM}, this check can be done in $O(1)$ time with the $\IPM_S$ data structure. 
If $\root$ is not a substring of $\run$, then once again $\sub$ has no occurrences in $\run$ and so $\C=\varnothing$. 

From now on, we assume that $|\run|\ge 3\rho$ and $\run$ contains an occurrence of $\root$.
Then the algorithm computes, in $O(1)$ time by \cref{obs:dldr}, the parameters of the representation $\run = \root^{[d_\ell^\run;q_\run;d_r^\run]}$.
One has $q_\run\ge 2$ since $|\run|\ge 3\rho$.
We recall that $\sub=\root^{[d_\ell;q_{\ell,r};d_r]}$.

Note that $\sub$ covers index $i$ of $S$ with an occurrence contained in $\run$ if and only if it covers the index $j=i-a_\run + 1$ \emph{of the string $\run$}.
In order to build the set $\C$, we describe, in \cref{clm:subinrun}, a set of necessary and sufficient conditions for $\sub$ to cover an index $j$ of $\run$. 
We denote $b_\ell=[d_\ell > d_\ell^\run]$, $b_r=[d_r > d_r^\run]$ (Iverson bracket notation).
We need two auxiliary claims.

\begin{claim}\label{c:startingandendinginrun}
   Let $\sub$ occur in $\run$. 
   If $x$ is the starting index of its leftmost occurrence and $y$ is the ending index of its rightmost occurrence, then $x=d_\ell^\run - d_\ell + 1+ b_\ell\rho$, $y= d_\ell^\run +(q_\run-b_r) \rho + d_r$, and $[x..y]$ is exactly the set of indices covered by $\sub$ in $\run$.
\end{claim}

\begin{figure}[!htb]
    \centering
    \includegraphics[scale=0.9, trim = 33 665 140 35, clip]{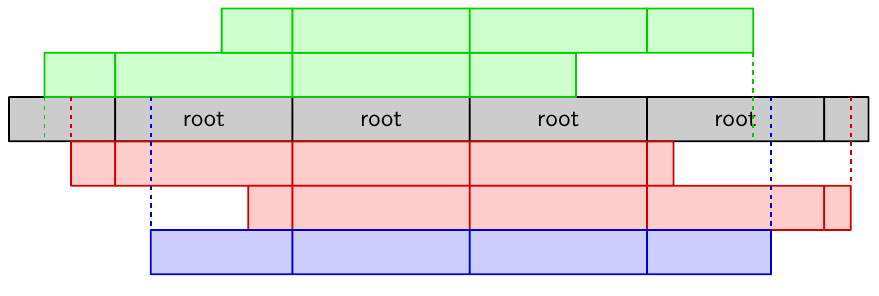}
    \caption{Occurrences of $\sub$ in $\run$ (\cref{clm:subinrun}).
    Grey strip is $\run$, color strips indicate occurrences of $\sub$ (one color for one substring $\sub$).
    The substrings drawn red, green, and blue realize, respectively, conditions~1, 3, and 4 of \cref{clm:subinrun}.
    Dash lines show ranges covered by $\sub$ in each case.}
    \label{fig:subinrin}
\end{figure}

\begin{claimproof}
    We prove the formula for $x$ as the argument for  $y$ is similar.
    Since $\run$ is $\rho$-periodic, we have $x\in[1..\rho]$ (otherwise, there is another occurrence at position $x-\rho$).
    Since $\root$ occurs in $\sub$ at position $d_\ell+1$, $\run$ has a matching occurrence of $\root$ at position $d_\ell+x\in[d_\ell+1..d_\ell+\rho]$.
    The occurrences of $\root$ in $\run$ are at positions $d_\ell^\run{+}1,d_\ell^\run{+}\rho{+}1,\ldots$, so exactly one of them starts in $[d_\ell+1 .. d_\ell + \rho]$.
    If $d_\ell \le d_\ell^\run$, one has $d_\ell^\run+1\in[d_\ell+1..d_\ell+\rho]$.
    Therefore, we have $d_\ell^\run+1=d_\ell+x$, implying $x=d_\ell^\run-d_\ell+1+0\cdot\rho$ as required.
    Similarly, if $d_\ell> d_\ell^\run$, one has $d_\ell^\run+\rho+1\in[d_\ell+1..d_\ell+\rho]$.
    Then $d_\ell^\run+\rho+1=d_\ell+x$, which implies $x=d_\ell^\run-d_\ell+1+1\cdot\rho$ as required.

    Since $\run$ is $\rho$-periodic, the positions of any two consecutive occurrences of $\sub$ in $\run$ differ by $\rho$ (see \cref{fig:subinrin}).
    As $|\sub|>\rho$, the indices in $\run$ covered by occurrences of $\sub$ form a single range from the first index of the leftmost occurrence (i.e., $x$) to the last index of the rightmost occurrence (i.e. $y$).
\end{claimproof}

\begin{claim}\label{c:substringcondition}
    The string $\sub$ occurs in $\run$ if and only if $q_{\ell,r} \le q_\run-b_\ell-b_r$.
\end{claim}
\begin{claimproof}
    Let $\sub$ occur in $\run$. 
    By \cref{c:startingandendinginrun}, its leftmost occurrence is at $x = d^\run_\ell - d_\ell + 1 + b_\ell\rho $ and its rightmost occurrence ends at $y=d_\ell^{\run} + (q_\run-b_r)\rho +d_r$.
    Clearly, we have the inequality $y-x +1 \ge |\sub| = d_\ell + q_{\ell,r}\rho + d_r$, which is equivalent to $q_{\ell,r} \le q_\run - b_\ell - b_r$.

    For the converse, we assume that this inequality holds and show that $\sub$ occurs in $\run$ at position $x=d_\ell^\run - d_\ell + 1 + b_\ell\rho$.
    Since $\sub$ and $\run$ are both $\rho$-periodic and share the substring $\root$ of length $\rho$, it suffices to prove that $|\run[x..|\run|]|\ge |\sub|$.
    Observing that $b_r\rho\ge d_r-d_r^\run$, we obtain
    \begin{multline*}
        |\run[x..|\run|]| = q_\run\rho+d_\ell^\run+d_r^\run - x +1 =
        (q_\run-b_\ell)\rho + d_\ell+d_r^\run\\
        \ge (q_\run-b_\ell-b_r)\rho + d_\ell+d_r
        \ge q_{\ell,r}\rho + d_\ell+d_r = |\sub|,
    \end{multline*}
    as required.
\end{claimproof}
\begin{claim} \label{clm:subinrun}
    The string $\sub=\root^{[d_\ell;q_{\ell,r};d_r]}$ covers index $j$ in $\run = \root^{[d_\ell^\run;q_\run;d_r^\run]}$ if and only if one of the following mutually exclusive conditions holds:
\begin{enumerate}
    \item $d_\ell \le d_\ell^\run$, $d_r \le d_r^\run$, $q_{\ell,r} \le q_\run\phantom{\,-\,1}$, and $j\in[d_\ell^\run-d_\ell+1\:..\:d_\ell^\run+q_\run\rho+d_r]$;
    \item $d_\ell > d_\ell^\run$, $d_r \le d_r^\run$, $q_{\ell,r} \le q_\run -1$, and $j\in[d_\ell^\run-d_\ell+1+\rho\:..\:d_\ell^\run+q_\run\rho+d_r]$;
    \item $d_\ell \le d_\ell^\run$, $d_r > d_r^\run$, $q_{\ell,r} \le q_\run -1$, and $j\in[d_\ell^\run-d_\ell+1\:..\:d_\ell^\run+(q_\run-1)\rho+d_r]$;
    \item $d_\ell > d_\ell^\run$, $d_r > d_r^\run$, $q_{\ell,r} \le q_\run -2$, and $j\in[d_\ell^\run-d_\ell+1+\rho\:..\:d_\ell^\run+(q_\run-1)\rho+d_r]$.
\end{enumerate}
\end{claim}

\begin{claimproof}
    If $\sub$ covers index $j$, then $\sub$ occurs in $\run$.
    Hence \cref{c:substringcondition} implies the inequalities and \cref{c:startingandendinginrun} implies the range for each of conditions 1--4.

    For the converse, if one of the conditions 1--4 is true, then $\sub$ indeed occurs in $\run$ according to \cref{c:substringcondition}.
    Then again \cref{c:startingandendinginrun} implies the range of indices covered by occurrences of $\sub$.
    As $j$ belongs to this range, it is covered.
\end{claimproof}

The algorithm builds the set $\C$ by running through conditions 1--4 of \cref{clm:subinrun}. 
If $j=i-a_\run+1$ belongs to the range from a condition, the algorithm adds to $\C$ the cuboid defined by the inequalities listed in this condition; otherwise, it does nothing.
The cuboids for the conditions 1, 2, 3, and 4 are, respectively, 
$[1..d_\ell^\run] \times [1..d_r^\run]\times [1..q_\run]$;
$[d_\ell^\run{+}1..\rho] \times [1..d_r^\run]\times [1..q_\run{-}1]$;
$[1..d_\ell^\run] \times [d_r^\run{+}1..\rho]\times [1..q_\run{-}1]$;
$[d_\ell^\run{+}1..\rho] \times [d_r^\run{+}1..\rho]\times [1..q_\run{-}2]$.

\para{Correctness.} Let $\sub$ cover $i$ with an occurrence contained in $\run$.
Then $\sub$ covers the index $j=i-a_\run+1$ in $\run$.
By \cref{clm:subinrun}, the triple $(d_\ell,d_r,q_{\ell,r})$ satisfies one of conditions 1--4, say, condition~N.
In particular, $j$ belongs to the interval of condition~N.
Then the algorithm built a cuboid $C$ from the inequalities of condition~N such that $(d_\ell,d_r,q_{\ell,r})\in C$.
Conversely, if $(d_\ell,d_r,q_{\ell,r})\in C$, where $C$ was built from condition~N of \cref{clm:subinrun}, then $j$ belongs to the interval of condition~N.
Hence condition~N holds; by \cref{clm:subinrun}, $\sub$ covers the index $j$ in $\run$, and thus covers $i$ in $S$.

As the time complexity is straightforward,
\cref{lem:runinrun}, and then
\cref{lem:rangecoverperRuntime}, is proved.

\section{2-Covers Oracle}\label{sec:oracle}
In this section, we present a solution to the \textsf{2-cover\_Oracle} problem (\cref{thm:oracle}).
The preliminary part of the solution is common to all three problems.

Every 2-cover of $S$ contains a prefix and a suffix of $S$.
Respectively, each 2-cover has one of two types (see~\cite{RS20}): a prefix-suffix 2-cover (\emph{ps-cover}) consists of a prefix of $S$ and a suffix of $S$ while in a border-substring 2-cover (\emph{bs-cover}) one string is a border of $S$.
We process these two cases separately.

Let $(U_1,U_2)$ be a pair of substrings.
\cref{lem:rangecoveraperRuntime,lem:rangecoverperRuntime} allow us to express each predicate ``$U_j$ covers index $i$'' as $p_j\in\R_j^i$, where $p_j$ is a point and $\R_j^i$ is a set of $O(1)$ ranges in $d_j$ dimensions.
Then the predicate ``$(U_1,U_2)$ is a 2-cover'' is expressed by the 2CNF formula $\bigwedge_{i=1}^n(p_1\in\R_1^i \vee p_2\in\R_2^i)$.
We answer the instances of this predicate with a new data structure based on rectangle stabbing (\cref{lem:stab}). The lemma below is proven in \cref{app:2cnf}.
\begin{restatable}[2CNF Range Data Structure]{lemma}{lemtwocnfds}
\label{lem:2CNFDS}
    Let $d_\ell,d_r$ be integer constants and let $\Pairs=\{(\L_1,\R_1),(\L_2,\R_2),\ldots,(\L_n,\R_n)\}$ be a set of pairs such that for every $i\in[n]$, $\L_i$ is a set of $O(1)$ $d_\ell$-dimensional orthogonal ranges and $\R_i$ is a set of $O(1)$ $d_r$-dimensional orthogonal ranges.
    The set $\Pairs$ can be prepossessed in $O(n\log^{d_\ell+d_r-1}n)$ time to a data structure that supports the following query in $O(\log^{d_\ell+d_r-1}n)$ time:
    \begin{itemize}
        \item $\mathsf{query}(p_\ell,p_r)$: for a $d_\ell$-dimensional point $p_\ell$ and a $d_r$-dimensional point $p_r$, decide if for every $i\in [n]$ either $p_\ell\in \L_i$ or $p_r\in \R_i$.
    \end{itemize}
\end{restatable}

As \cref{lem:rangecoveraperRuntime} refers to particular ranges and \cref{lem:rangecoverperRuntime} refers to particular periods, we partition substrings into groups and build a separate $2\CNF$ data structure for each pair of groups.
For prefixes, suffixes, and borders, we have $O(\log n)$ periods (\cref{lem:fewperiods}) and thus $O(\log n)$ groups of highly periodic prefixes (suffixes, borders).
The remaining prefixes (suffixes) form $O(\log n)$ groups associated with length ranges $[1.5^k..1.5^{k+1}]$ for some $k$.
There are $O(\log n)$ remaining borders (\cref{lem:fewborders}), so each of them forms a separate group.
For each group of borders we choose a fixed position $f$. 
Highly periodic substrings containing $f$ form $O(\log n)$ groups (\cref{lem:runscover}); the other are grouped according to $O(\log n)$ length ranges.
Therefore, in total we build $O(\log^2 n)$ $2\CNF$ data structures for ps-covers and bs-covers.

\para{Effective dimension.}
A direct implementation of \cref{lem:rangecoveraperRuntime,lem:rangecoverperRuntime} leads to the $2\CNF$ structures of dimension 4 to 6.
Let us show how to lower the dimension.
For any group $\pref$ of prefixes we take $f=1$. 
Then in \cref{lem:rangecoveraperRuntime} all points have the form $(0,r)$.
So we have \emph{fixed} first coordinate and \emph{variable} second coordinate.
In \cref{lem:rangecoverperRuntime}, one has $\root=S[1..\rho]$, and thus all points have the form $(0,d_r,q_{r})$ with two variable coordinates.
For groups of suffixes we take $f=n$ and symmetrically get the points of the form $(\ell,0)$ or $(d_\ell,0,q_{\ell})$.
Since borders are simultaneously prefixes and suffixes, we get two fixed coordinates in the corresponding points. (Assuming $f=1$, a group consisting of a single border $U$, has the point $(0,|U|)$; the group $\bor$ of highly $\rho$-periodic borders has the points $(0,d_r, q)$, where the remainder $d_r=|U|\bmod \rho$ is the same for all $U\in \bor$.)
Finally, for general substrings all coordinates are variable.
The \emph{effective dimension} of a point is the number of its variable coordinates.
Given a pair of groups of substrings, where the first (second) group has points of \emph{effective} dimension $d_1$ (respectively, $d_2$), we construct for them the $2\CNF$ structure of dimension $d_1+d_2$.
In order to do this, we replace each involved range with its projection onto variable coordinates.

\para{Building an oracle.}
Given a group $\pref$ of not highly periodic (resp., highly periodic) prefixes, we apply \cref{lem:rangecoveraperRuntime} (resp.,  \cref{lem:rangecoverperRuntime}) for every $i\in[n]$.
Let $\L_1,\ldots \L_n$ be the projections of the obtained ranges onto variable coordinates.
A group $\suff$ of suffixes is processed in the same way, resulting in the ranges $\R_1,\ldots, \R_n$.
Then we apply \cref{lem:2CNFDS}, constructing the $2\CNF$ structure over the set $\Pairs=\{(\L_1,\R_1),\ldots,(\L_n,\R_n)\}$.
This $2\CNF$ thus represents the set $\pref\times \suff$ of pairs of substrings.
We also memorize the values of fixed coordinates.
Iterating over all pairs of prefix and suffix groups, we obtain the ps-cover part of the oracle.

Given a group $\bor$ of borders, we first determine the reference position $f_\bor$ for groups of substrings and store it.
If $\bor=\{U\}$, we use \cref{lem:PM} to find all occurrences of $U$ in $S$ and choose $f_\bor$ to be any position not covered by $U$; if there is no such position, i.e., if $U$ is a 1-cover, we set $f_\bor=\infty$.
If $\bor$ is a group of highly $\rho$-periodic borders, we run a binary search on it, finding the shortest border $U$ that is not a 1-cover.
Then we choose a position $f_\bor$ not covered by $U$.
Additionally, we store $|U|$.
If $U$ does not exist, we set $f_\bor=\infty$.
After determining $f_\bor$, and only if it is finite, we build a $2\CNF$ structure similar to the prefix-suffix case.
Iterating over all pairs of border and substring groups (for the latter, we fix $f=f_\bor$), we obtain the bs-cover part of the oracle.

The time complexity of the construction is dominated by building $O(\log^2n)$ $2\CNF$ structures, each of dimension at most 4 (in the case where both groups consist of highly periodic strings).
By \cref{lem:2CNFDS}, we get the required $O(n\log^5n)$ time bound.

\para{Querying an oracle.}
Given a pair $(U_1,U_2)$ of substrings of $S$, the oracle decides with $\LCP$ queries, to which of the cases (prefix-suffix, border-substring, both, or neither) this pair can be attributed, and proceed accordingly.
For prefix, suffix, or border, the oracle finds its group deciding high periodicity with a query to $\IPM_S$ (\cref{lem:IPM}).
In the prefix-suffix case the oracle then create points for $U_1$ and $U_2$, ``trim'' them by dropping fixed coordinates, and query with this pair of trimmed points the $2\CNF$ structure built for the set $\pref\times\suff$, where the groups $\pref$ and $\suff$ contain $U_1$ and $U_2$ respectively.

Consider the border-substring case (let $U_1$ be the border). 
After determining the group $\bor$ of $U_1$, we check $f_\bor$.
If $f_\bor=\infty$, the oracle returns True since $U_1$ is a 1-cover.
The same applies for the case where $f_\bor$ is finite, $\bor$ is highly periodic, and $U_1$ is shorter than the saved length $|U|$.
Otherwise, we create the point for $U_1$, ``trim'' it by dropping fixed coordinates, and create the point for $U_2$ using $f=f_\bor$.
Then we query with the obtained pair of points the  $2\CNF$ structure built for the pair $\bor\times\sub_f$, where the group $\sub_f$ contains $U_2$.

Finally, the oracle returns True if it met a condition for ``True'' in the border-substring case or if some query made to a $2\CNF$ structure returned True.
Otherwise, the oracle returns False. 
The query time is dominated by $O(1)$ queries to  $2\CNF$ structures, each of dimension at most 4. By \cref{lem:2CNFDS}, we get the required $O(\log^3n)$ time bound.

As a result, we proved \cref{thm:oracle}.
The omitted details can be found in \cref{app:oracle}.

\section{Reporting 2-Covers}\label{sec:report}

A possible, but in general highly inefficient, approach to the \textsf{All\_2-covers} and  \textsf{Shortest\_2-cover} problems is to construct the oracle of \cref{thm:oracle} and query it with every pair (of substrings) that can be in the answer.
In this section, we describe our approach to achieve near-linear running time.
In a high level, we build a fast reporting procedure for ``simple'' cases and use its answer to determine the rest of the output with a small number of oracle queries.
To rule out the trivial situation, we assume that all 2-covers containing a 1-cover are already reported just by listing the 1-covers.

We call a 2-cover $(X,Y)$ \emph{core} if both substrings $X$ and $Y$ are not highly periodic.
This means that the $2\CNF$ structure for their groups is built by using only \cref{lem:rangecoveraperRuntime}, and thus is 2-dimensional.
In particular, a core cover is associated with a 2-dimensional point $(x,y)$.
Note that 2-dimensional $2\CNF$ structures represent all core 2-covers (and may represent some non-core 2-covers as certain highly periodic substrings pass the restriction on periods in statement~1 of \cref{lem:rangecoveraperRuntime}).
The shortest 2-cover is core in view of \cref{lem:removePeriod}.

On the ground level, a $d$-dimensional $2\CNF$ structure stores a set $\R$ of $O(n)$ $d$-dimensional ranges and checks whether a $d$-dimensional point, sent as a query, is outside all rectangles.
Such a view inspires the following definition for the case $d=2$ we are interested in. 

\begin{restatable}[Free Point]{definition}{deffreepoint}
\label{def:freepoint}
Let $\R$ be a set of rectangles with corners in $[n]^2$.
A point $p\in [n]^2$ is $\R$-free if $p\notin R$ for every $R\in \R$.
\end{restatable}

The following lemma is crucial.
For the full proof see \cref{app:FreePoint}.

\begin{restatable}[Free Points Reporting]{lemma}{lemAllFreePoints}
\label{lem:AllFreePoints}
Let $\R$ be a set consisting of $\Theta(n)$ rectangles with corners in $[n]^2$.
There is an algorithm that reports, for the input $\R$,
\begin{itemize}
    \item all $\R$-free points in $O(n\log^2 n+\Output\cdot \log n)$ time, or
    \item for each $y\in[n]$, the $\R$-free point $(x,y)$ with minimal $x$ (if any) in $O(n\log^2 n)$ time, or
    \item for an additional input $m\in [n]$, all $\R$-free points $(x,y)$ with $x+y \le m$ in time $O(n\log^2 n + \Output \cdot \log(n))$.
\end{itemize}
\end{restatable}

The main point of the algorithm of \cref{lem:AllFreePoints} is an efficient reduction of the 2-dimensional problem to its 1-dimensional analog: for each $y\in[n]$, report all points in the complement of a union of $x$-ranges, corresponding to this $y$.
The algorithm stores the total of $O(n\log n)$ $x$-ranges in an auxiliary tree $\cT$ and associates with each node of $\cT$ a version of the main structure $\cD$, which is a variant of persistent lazy segment tree \cite{RuSh17}.

To present a solution to the \textsf{Shortest\_2-cover} problem, we need the following claim.

\begin{claim} \label{clm:pointtocover}
    If a point $(x,y)$ is associated with a core 2-cover $(X,Y)$, then $|X|+|Y|=x+y$ in the prefix-suffix case and $|X|+|Y|=x+y+1+b$ in the border-substring case with the border of length $b$.
\end{claim}
\begin{claimproof}
    In the prefix-suffix case, $X=S[1..x]$ and $Y=S[n-y+1..n]$.
    In the border-substring case, $X=S[1..b]$ and $Y=S[f-x..f+y]$ for some position $f$.
    The claim follows.
\end{claimproof}

A solution to \textsf{Shortest\_2-cover} is as follows.
We build all 2-dimensional $2\CNF$ structures (\cref{lem:2CNFDS}).
For the set $\R$ of each structure, we run the algorithm of \cref{lem:AllFreePoints} with the second option and choose the point $(x,y)$ with the minimum sum of coordinates.
From this pair we restore the corresponding 2-cover $(X,Y)$.
Due to \cref{clm:pointtocover}, $(X,Y)$ has the minimum length among all $2$-covers corresponding to this $2\CNF$ structure.
After processing all sets $\R$, we return the $2$-cover of minimum length among those found.

Since each of $O(\log^2n)$ sets $\R$ is computed in $O(n\log n)$ time (\cref{lem:2CNFDS}) and processed in $O(n\log^2 n)$ time (\cref{lem:AllFreePoints}), the time complexity is $O(n\log^4 n)$, as required.
\cref{thm:shortest} is proved.

\subsection{Report All 2-Covers Of Bounded Length}
In this section, we overview the proof of \cref{thm:reporting},
focusing on reporting all ps-covers with length bounded by $m$. 
The process of reporting all bs-covers is similar, and the full details are presented in \cref{app:report}.

As a preliminary step, the algorithm constructs the $2$-cover oracle of \cref{thm:oracle}.
The first main step is similar to the proof of \cref{thm:shortest}: the algorithm computes the set $\C_m$ of all core $2$-covers of length at most $m$ 
using the third variant of \cref{lem:AllFreePoints}.

The remaining task is to report all highly periodic $2$-covers with length bounded by $m$.
This is achieved via an extending procedure based on the following observation.

\begin{observation}\label{obs:xytrim}
Let $(X,Y)$ be a ps-cover of $S$.
Let $X_\trim = X$ if $X$ is not highly periodic, or $X_\trim = X[1..2\rho+|X| \bmod \rho]$ if $X$ is highly $\rho$-periodic.
Similarly, let $Y_\trim = Y$ if $Y$ is not highly periodic, or $Y_\trim = Y[|Y| - 2\rho - |Y|\bmod \rho +1..|Y|]$ if $Y$ is highly $\rho$-periodic.     

The pair $(X_\trim,Y_\trim)$ is a core $2$-cover of $S$.
\end{observation}
The fact that $(X_\trim,Y_\trim)$ is a $2$-cover follows directly from \cref{lem:removePeriod}, and the fact that it is core is immediate: if $X$ is not short periodic, so is $X_\trim = X$, and if $X$ is highly periodic, $X_\trim$ is its short $\rho$-periodic prefix with the same length modulo $\rho$. 
The same analysis applies to $Y$ and $Y_\trim$.
Clearly, if the length of $(X,Y)$ is bounded by $m$, we have $(X_\trim,Y_\trim) \in \C_m$.

We proceed to describe how to exploit \cref{obs:xytrim} to report all prefix-suffix $2$-covers with length bounded by $m$.
We process every pair $(X,Y) \in \C_m$. 
When processing $(X,Y)$, we wish to report all $2$-covers $(X',Y')\notin \C_m$ with $|X'|+|Y'| \le m$, $X=X'_\trim$ and $Y=Y'_\trim$.
Note that such $(X',Y')$ may exist only if $X$ or $Y$ is short periodic.
\cref{obs:xytrim} directly implies that all non-core ps-covers of length at most $m$ are reported in this way. 
Assume that $X=S[1..2\rho + d]$ is short $\rho$-periodic and $Y$ is aperiodic.
We initialize an iterator $q=3$, and check if the following conditions hold:
\begin{enumerate}
    \item $q\cdot \rho + d  + |Y| \le m$.
    \item $X_q = S[1..q\cdot \rho+d]$ is $\rho$ periodic.
    \item the pair $(X_q,Y)$ is a $2$-cover.
\end{enumerate}
The second condition is checked using \cref{lem:IPM}, and the third condition is checked via a query to the $2$-cover oracle.
If all three conditions are true, the algorithm reports $(X_q,Y)$ as a $2$-cover, assigns $q \leftarrow q+1$ and checks the conditions again.
Otherwise, the algorithm halts.
It is easy to see that exactly all $2$-covers $(X',Y')$ such that $(X,Y)$ is the trimmed version of $(X',Y')$ are reported this way.

The time complexity is dominated by the queries to the oracle.
Each query that returns True can be charged on the reported $2$-cover. Every query that returns False can be charged on the original pair $(X,Y)$, as a 'False' response terminates the algorithm.
It follows that the total running time on every $(X,Y) \in \C_m$ is $O(\Output \cdot \log^3 n)$ as required.

The case in which $X$ is aperiodic and $Y$ is short periodic is completely symmetrical, and the case in which both are short periodic is treated in a 'nested loop' fashion: only $X$ is extended until breaking one of the conditions, and then the process repeats with $Y$ extended by one period, two periods, and so on.
In \cref{app:report}, we show how to appropriately execute this nested loop in a manner that guaranteed the desired running time.

\bibliography{covers}

\appendix
\section{Figures}

\section{Missing Proofs}\label{app:pre}

The following two well-known lemmas are useful for the proofs in this section.
\begin{lemma}[Fine and Wilf Theorem~\cite{FiWi65}]\label{lem:FW}
    Every string having periods $p,q$ and length at least $p+q-\gcd(p,q)$, has period $\gcd(p,q)$.
\end{lemma}

\begin{lemma}[Rephrased Three Squares Lemma \cite{CrRy95}]\label{lem:3squares}
    If a string $S$ has periodic prefixes $X$, $Y$, and $Z$ such that $\per(X)<\per(Y)<\per(Z)$, then $\per(Z)\ge\per(X)+\per(Y)$.
\end{lemma}

We complete here all missing proofs of \cref{sec:prelim}.

\begin{proof}[Proof of \cref{lem:fewperiods}]
    Let $p_1<p_2<\cdots<p_\ell$ be all  periods of periodic prefixes of the string $S[1..n]$.
    Consider three consecutive values $p_i,p_{i+1},p_{i+2}$ from this list.
    By \cref{lem:3squares}, $p_{i+2} > 2p_i$.
    Since $p_\ell \le n$, we immediately get $\ell =O(\log n)$ as required.
\end{proof}

\begin{proof}[Proof of \cref{lem:fewborders}]
    Let $b_1<b_2<\cdots< b_\ell$ be all different lengths of \emph{aperiodic} borders of the string $S[1..n]$.
    Since for every $i \in [1..\ell-1]$ the string $S[1..b_i]$ is a border of $S[1..b_{i+1}]$, $b_{i+1}-b_i$ is a period of $S[1..b_{i+1}]$.
    Then $(b_{i+1}-b_i)>b_{i+1}/2$ because of aperiodicity.
    Hence $b_{i+1}>2b_i$.
    Since $b_\ell \le n$, we get $\ell = O(\log n)$.
    
    For \emph{short periodic} borders, a similar argument gives the inequalities $b_{i+1}> 1.5 b_i$, which also bound the number of borders by $O(\log n)$.
\end{proof}

\begin{proof}[Proof of \cref{lem:removePeriod}]
    By definition of period, $X[1..|X|-\rho]=Y$ implies $X[\rho+1..|X|]=Y$.
    Since $\rho \le \frac{|X|}{2}$, the string $Y$ covers all indices in $X$ with these two occurrences.
    It follows that $Y$ covers every index $i$ in $S$ that is covered by $X$.
\end{proof}

\begin{proof}[Proof of \cref{lem:runscover}]
    Clearly, an occurrence of a substring of length $\rho$ is contained in at most one $\rho$-periodic run.
    Then the first statement of the lemma stems from the fact that a $\rho$-periodic run covering $i$ contains either $S[i-\rho+1..i]$ or $S[i..i+\rho-1]$ (or both).
    For the second statement, some work is needed.

    With respect to the index $i$, we call a run \emph{left} (resp., \emph{right}, \emph{central}) if it contains the substring $S[i-2\rho+1..i]$ (resp., $S[i..i+2\rho-1]$, $S[i-\rho..i+\rho]$), where $\rho$ is the period of the run.
    Every highly-periodic run covering $i$ possesses at least one of these three properties.
    By \cref{lem:3squares}, the number of right runs is $O(\log n)$; the dual of \cref{lem:3squares} implies the same result for left runs.
    Now let $A$ and $B$ be two highly-periodic central runs, with $\rho=\per(A)$, $\rho'=\per(B)$. 
    If neither of $A$, $B$ contains the other one, then their overlap is of length at least $\rho+\rho'$ (otherwise, one of them is not central).
    This overlap is a string with periods $\rho$ and $\rho'$, and then has period $\gcd(\rho,\rho')$ by \cref{lem:FW}.
    Hence $A$ and $B$ are substrings of some $\gcd(\rho,\rho')$-periodic run, contradicting our assumption that $A$ and $B$ are runs.
    Therefore, one of $A,B$ contains the other.
    Without loss of generality, $B$ contains $A$.
    Then $\rho\ne \rho'$.
    The string $A$ has periods $\rho$ and $\rho'$ but no period $\gcd(\rho,\rho')$ (otherwise it is not a $\rho$-periodic run).
    Then $|A|<\rho+\rho'$ by \cref{lem:FW}; as $|A|\ge 3\rho$, we have $\rho'>2\rho$. 
    This immediately implies the $O(\log n)$ upper bound on the number of central runs.
    The lemma now follows.
\end{proof}

\begin{proof}[Proof of \cref{lem:uniquerun}]
First notice that since $S[i..j]$ is $\rho$ periodic, there is at least one $\rho$-periodic run containing $S[i..j]$.
Assume to the contrary that there are two different $\rho$-periodic runs $A$ and $B$ containing $S[i..j]$.
Since $A$ and $B$ overlap by a range of length $\rho$, the entire range covered by $A$ and $B$ is $\rho$-periodic, which contradicts the maximality of $A$ and $B$.
\end{proof}

\begin{proof}[Proof of \cref{lem:runextaper}]
    Assume to the contrary that $S[x..y+1]$ is $\rho'$-periodic for some $\rho'$.
    If $\rho'$ is divisible by $\rho$, we have $S[y+1] = S[y-\rho'+1] = S[y - \rho+1]$ where the second equality is due to the $\per(S[x..y]) = \rho$.
    Then $S[x..y+1]$ is $p$-periodic, contradicting the condition of the lemma.
    If $\rho'$ is not divisible by $\rho$, by \cref{lem:FW} we have that $\per(S[x..y])$ is smaller than $\rho$, contradicting the definition of $\rho$-periodic.
\end{proof}

\begin{proof}[Proof of \cref{lem:inverseofhypercubes}]
    We present a proof for a set of $2$-dimensional ranges, i.e., rectangles.
    This proof can be easily generalized for any constant dimension.
    Consider the infinite extensions of every side of a rectangle $R \in \R$ to both directions.
    Namely, for a rectangle $R =[x_1..x_2]\times[y_1..y_2]$, the infinite extensions of the sides of $R$ are the lines $x = x_1$, $x=x_2$, $y = y_1$ and $y=y_2$.
    Clearly, the infinite extensions of the sides of all rectangles in $\R$ partition the plane into $O(|\R|^2) = O(1)$ rectangles.
    Every rectangle in this partition is either contained in a rectangle of $R$, or is disjoint from all rectangles of $R$.
    The set of rectangles in the partition disjoint from all rectangles of $R$ satisfies the claim.
    Moreover, this set can be computed in $O(1)$ time straightforwardly.
\end{proof}

\section{2CNF Data Structure}\label{app:2cnf}

In this section we prove \cref{lem:2CNFDS}.
We use the following auxiliary lemma to manipulate $d$-dimensional ranges.

\begin{lemma}[Inverse of Ranges]\label{lem:inverseofhypercubes}
    Let $\R$ be a set of $O(1)$ $d$-dimensional ranges, where $d$ is an integer constant.
    There is a set $\overline{\R}$ of $O(1)$ $d$-dimensional ranges such that $\bigcup_{R\in \overline{\R}}R = [-\infty .. \infty]^d \setminus \bigcup_{R \in \R}R$.
    Moreover, the set $\overline{R}$ can be computed in $O(1)$ time given $\R$.
\end{lemma}

\lemtwocnfds*
\begin{proof}
Let $d=d_\ell+d_r$.
We build a set $\B$ of $d$-dimensional ranges, processing each pair $(\L_i,\R_i)$ as follows.
We start with the set $B_i$ of $d$-dimensional ranges defined by
$$
B_i=\{ L\times[-\infty..\infty]^{d_r}\mid L\in \L_i\}\cup 
\{ [-\infty..\infty]^{d_\ell}\times R\mid R\in \R_i\}.
$$
As $|B_i|=|\L_i|+|\R_i|=O(1)$, we apply \cref{lem:inverseofhypercubes} to get, in $O(1)$ time, its inverse set of ranges $\overline{B_i}$, which is also of size $O(1)$.
Now let $\B=\bigcup_{i\in[n]}\overline{B_i}$.

We preprocess the set $\B$ into a range stabbing data structure (\cref{lem:stab}).
To answer $\mathsf{query}(p_\ell,p_r)$, where $p_\ell=(x_1,\ldots,x_{d_\ell})$ and $p_r=(y_1,\ldots,y_{d_r})$, we perform the Existence query to this structure with the $d$-dimensional point $p=(x_1,\ldots,x_{d_\ell},y_1,\ldots,y_{d_r})$ and report the \textit{negation} of the obtained answer.
The required time complexities follow from \cref{lem:stab}.

\para{Correctness.}
We need to prove that $p\notin \bigcup_{B\in \B} B$ if and only if for every $i\in[n]$ either $p_\ell\in L$ for some $L\in \L_i$ or $p_r\in R$ for some $R\in \R_i$.

Let $p\notin \bigcup_{B\in \B}B$ and let $i\in[n]$.
By definition of $\B$, $p\notin \bigcup_{B\in \overline{B_i}}B$.
Then $p\in \bigcup_{B\in B_i}B$.
By definition of $B_i$, this implies that $p_\ell\in L$ for some $L\in \L_i$ of $p_r\in R$ for some $R\in \R_i$.

For the converse note that if $p_\ell\in L$ for some $L\in \L_i$ of $p_r\in R$ for some $R\in \R_i$, then $p\in B$ for some $B\in B_i$ and hence $p\notin B'$ for all $B'\in \overline{B_i}$.
If this is the case for all $i$, then $p\notin \bigcup_{B\in \B}B$ by definition.
\end{proof}

\section{2-Covers Oracle}\label{app:oracle}
In this section, we provide a detailed proof of \cref{thm:oracle}.
We distinguish between two types of $2$-covers. 
A $2$-cover $(C_1,C_2)$ is a prefix-suffix cover (ps-cover) if $C_1$ is a prefix of $S$ and $C_2$ is a suffix of $S$ (or vice versa).
A $2$-cover is a border-substring cover (bs-cover) if $C_1$ is a border of $S$ and $C_2$ is a substring of $S$ (or vice versa).
Note that every $2$-cover can be classified into one of these two types.
More specifically, the oracle queries each data structure twice: once for $(C_1,C_2)$ and once for $(C_2,C_1)$.

Our data structure consists of two independent data structures, one for checking if $(C_1,C_2)$ is a prefix-suffix cover, and one for checking if it is a border-substring cover. 
When receiving a query, the data structure checks both options and answers accordingly.

\para{Common Preprocess.}
In both data structures, the following preprocess on $S$ is applied in addition to the preprocessing phase described in \cref{sec:ranges}.
The algorithm constructs an interval stabbing data structure $D^s_{hp}$ that contains an interval $[i..j]$ for every highly periodic run $S[i..j]$ in $S$ using \cref{lem:stab}.
This preprocess takes $O(n\log^2 n)$ time.

\subsection{Prefix-Suffix Oracle}\label{sec:psoracle}

In this section, we describe the oracle checking if $(U_1,U_2)=(S[1..p],S[x..y])$ is a ps-cover.

\para{Construction.}
We partition prefixes and suffixes into \emph{groups}.
The groups $\pper(\rho)$ and  $\sper(\rho)$ consist of all highly $\rho$-periodic prefixes (resp., suffixes).
The groups $\paper(k)$ and $\saper(k)$ consist of all not highly periodic prefixes (resp., suffixes) that have length in the range $[1.5^k..1.5^{k+1}]$.
Let
\begin{align*}
    \Pref &= \{ \paper(k)\mid 1.5^k \le n \} \cup \{\pper(\rho) \mid \rho \text{ is a prefix period} \},\\
    \Suf &= \{ \saper(k) \mid 1.5^k \le n \} \cup \{\sper(\rho) \mid \rho \text{ is a suffix period} \}.
\end{align*}
As the number of prefix/suffix periods is $O(\log n)$ by \cref{lem:fewperiods}, one has $|\Pref|,|\Suf|\in O(\log n)$.

The algorithm processes each pair $(\preftype,\sufftype) \in \Pref \times \Suf$ as follows.

\begin{itemize}
    \item If $\preftype = \paper(k)$, it uses, for every $i\in[n]$, \cref{lem:rangecoveraperRuntime} with parameters $f=1$, $i$, and $k$ to build a set of rectangles $\R_i$.
    From the construction given in \cref{clm:bsvalidrectangle} we see that each rectangle in $\R_i$ has the form $R=[0..0]\times[r_1..r_2]$, and every prefix in $\paper(k)$ is associated with a point of the form $p=(0,r)$.
    Then the inclusion $p\in R$ is decided solely from the second coordinates.
    We refer to the first coordinate as \emph{fixed} and we drop it to reduce the dimension.
    We thus define $\Pref_i$ to be the set of projections of all rectangles from $\R_i$ to the second coordinate.
    \item If $\preftype = \pper(\rho)$, it uses, for every $i\in[n]$, \cref{lem:rangecoverperRuntime} with parameters $f=1$, $i$, and $\rho$ to build a set of cuboids $\C_i$.
    Similar to the previous case we know that each $R\in \R_i$ has the first range $[0..0]$ and every prefix in $\pper(\rho)$ is associated with a point having zero first coordinate.
    Accordingly, we drop the fixed first coordinate and define $\Pref_i$ as the set of projections, to the last two coordinates, of all cuboids from $\C_i$.
\end{itemize}

The sets $\Suf_i$ are defined in a symmetric way from $\sufftype$; here we drop the fixed second coordinate.

Thus, the algorithm obtain the set $\Pairs=\{(\Pref_1,\Suf_1),(\Pref_2,\Suf_2),\ldots,(\Pref_n,\Suf_n)\}$ and builds a $2\CNF$ data structure $\CNF_{\preftype,\sufftype}$ of \cref{lem:2CNFDS} for the set $\Pairs$ with the following dimensions:
\[d_\ell=\begin{cases}
    1&\text{if }\preftype=\paper(k)\\
    2&\text{if }\preftype=\pper(\rho)
\end{cases}
\text{ and }
d_r=\begin{cases}
    1&\text{if }\sufftype=\saper(k)\\
    2&\text{if }\sufftype=\sper(\rho)
\end{cases}.
\]

\para{Query.}
For a query $(S[1..p],S[x..y])$ the algorithm first verifies that $S[x..y]$ is a suffix of $S$ by checking if $\LCP^R(y,n)\ge y-x+1$.
If it is not a suffix, the oracle answers False.
Otherwise, the algorithm queries the $\IPM_S$ data structure (\cref{lem:IPM}) with Periodic($S[1..p]$).
If the answer is ``aperiodic'' or a number $\rho>p/3$, the prefix is not highly periodic.
So the algorithm computes $k=\floor{\log_{1.5}p}$ and sets $\preftype = \paper(k)$ and $p_\ell=(p-1)$ (recall that the first coordinate is dropped!).
Otherwise, the prefix is highly $\rho$-periodic. 
In this case, the algorithm sets $\preftype = \pper(\rho)$ and $p_\ell=(d_r,q_r)$ for the integers $d_r\in[0..\rho-1]$ and $q_r$ such that $p=q_r\cdot\rho+d_r$.
Similarly, the algorithm queries $\IPM_S$ with Periodic($S[x..y]$).
If $S[x..y]$ is not highly periodic, the algorithm  computes $k=\floor{\log_{1.5}(y-x+1)}$ and sets $\sufftype = \saper(k)$ and $p_r=(y-x)$.
Otherwise, it is highly $\rho$-periodic, and the algorithm sets $\suff = \sper(\rho)$ and $p_r=(d_\ell,q_\ell)$ for the integers $d_\ell\in[0..\rho-1]$ and $q_\ell$ such that $y-x+1=q_\ell\cdot\rho+d_\ell$.
Finally, the algorithm queries the structure $\CNF_{\preftype,\sufftype}$ with the pair of points $(p_\ell,p_r)$ and returns the obtained answer.

\para{Complexity.}
The preprocessing phase takes $O(n\log^2n)$ time.
The periods of highly periodic prefixes and these prefixes themselves can be found in $O(n)$ time by $O(n)$ $\LCP$ queries $\LCP_S(S[1..n],S[\rho+1..n])$.
Highly periodic suffixes are operated in a symmetric way.
For each pair $(\preftype,\sufftype)$, the algorithm iterates every $i\in [n]$ and applies \cref{lem:rangecoveraperRuntime} or \cref{lem:rangecoverperRuntime}, which takes $O(n\log^2n)$ time.
Then, the algorithm constructs a 2$\CNF$ data structure of dimension $d_\ell + d_r \le 2+2=4$, which takes $O(n\log^3 n)$ time by \cref{lem:2CNFDS}.
Summing over $O(\log^2n)$ distinct pairs $(\preftype,\sufftype)$, we get the total running time of $O(n \log^{5}n)$. 

The query complexity is dominated by a single query to a $2\CNF$ data structure with $d_\ell +d_r \le 4$, which requires $O(\log^3n)$ time by \cref{lem:2CNFDS}.

\para{Correctness.}
By definition, $(S[1..p] , S[x..y])$  is a $2$-cover of $S$ if and only if for every $i$ we have $(S[1..p] , S[x..y])$ covers $i$.
Denote by $\Cover_i$ the event in which either $p_\ell$ is in some range of $\Pref_i$ or $p_r$ is in some range of $\Suf_i$ (where $\Suf_i$ and $\Pref_i$ refer to the sets used in the construction of $\CNF_{\preftype,\sufftype}$).
According to \cref{lem:rangecoveraperRuntime,lem:rangecoverperRuntime}, and our rule on dropping coordinates, an index $i$ is covered by $(S[1..p],S[x..y])$ if and only if $\Cover_i$ occurs.
It immediately follows that $(S[1..p],S[x..y])$ is a $2$-cover if and only if $\Cover_i$ occurs for every $i\in[n]$.
This is exactly the output of the $2\CNF$ query, as required.

\subsection{Border-Substring Oracle}\label{sec:bsoracle}

In this section, we describe the oracle checking if $(U_1,U_2)=(S[1..p],S[x..y])$ is a bs-cover.

\para{Construction.}
Similar to the prefix-suffix case, we partition borders into groups.
A group $\bor(\rho)$ consists of all highly $\rho$-periodic borders.
Each border that is not highly periodic, forms a separate group. 
By \cref{lem:fewperiods,lem:fewborders} the number of groups is $O(\log n)$.

With every group $\bor$ of borders, the algorithm associates the reference position $f_\bor$ for the groups of substrings that will be considered with this group of borders.
If $\bor=\{U\}$, the algorithm uses \cref{lem:PM} to find all occurrences of $U$ in $S$ and chooses $f_\bor$ to be any position not covered by $U$; if there is no such position, i.e., if $U$ is a 1-cover, it sets $f_\bor=\infty$.
If $\bor=\bor(\rho)$, the algorithm runs a binary search on $\bor$, finding the shortest border $U\in \bor(\rho)$ that is not a 1-cover.
This requires $O(n\log n)$ time as it uses \cref{lem:PM} $O(\log n)$ times.
Then the algorithm chooses a position $f_\bor$ not covered by $U$ and additionally stores $q_\bor=|U|$.
If $U$ does not exist, it sets $f_\bor=\infty$.

Let $f \in [n]$ be an index.
We define the sets $\subaper_f(k)$ and $\subper_f(\rho)$ in an analogous manner to the sets $\paper(k)$ and $\pper(\rho)$.
For an integer $k$, we define
\begin{align*}
\subaper_f(k) = \{S[f-\ell..f+r] \mid r,\ell \ge 0 \textit{ , } S[f-\ell..f+r] \textit{ is aperiodic, and } \ell+r-1 \in [1.5^{k} .. 1.5^{k+1}) \}
\end{align*}

For an integer $\rho$, we define
\begin{align*}
\subper_f(\rho) = \{ S[f-\ell..f+r] \mid r,\ell \ge 0 \textit{ and } S[f-\ell..f+r] \textit{ is highly } \rho \textit{-periodic}\}
\end{align*}

Note that every substring of $S$ that covers the index $f$ is either contained in $\subaper_f(k)$ for some value of $k$, or in $\subper_f(\rho)$ for some value of $\rho$.
The set $\S_f$ (analogous to $\Pref$ and $\Suf$) is defined as 
\begin{align*}
\S_f = \{\subaper_f(k) \mid 1.5^k \le n\} \cup \{ \subper_f(\rho)\mid \subper_f(\rho)\neq \emptyset \}
\end{align*}
Note that for a period $\rho$, every substring in $\subper_f(\rho)$ is contained in a highly $\rho$-periodic run that contains $f$.
It follows from \cref{lem:runscover}, that $|\S_f| \in O(\log n)$.

The algorithm iterates over all pairs $(\bor,\sub)$ where $\bor\in\B$ and $\sub\in \S_{f_\bor}$ (If $f_\bor = \infty$, we treat $\S_{f_\bor}$ as $\emptyset$).
Once again, there will be two fixed coordinates, but now both of them come from the border's side.
As the border is both a prefix (having fixed first coordinate) and a suffix (having fixed second coordinate), it has the first two coordinates fixed.

Consider processing of one pair $(\bor,\sub)$.
It results in creation of the set $\Pairs=\{(\B_1,\S_1),(\B_2,\S_2),\ldots,(\B_n,\S_n)\}$, which is then used to create a $2\CNF$ structure of \cref{lem:2CNFDS}.
If $\bor = \{U\}$, the algorithm applies \cref{lem:PM} to find the all indices in $S$ that are covered by $U$.
For each $i\in[n]$, we have a ``0-dimensional'' interval, which is just a boolean value indicating whether this position is covered by $U$ or not.
However, it is more convenient to represent it as a 1-dimensional interval, setting $\B_i=\{[-\infty .. \infty]\}$ if $i$ is covered and $\B_i =\varnothing$ if it is not.
If $\bor=\bor(\rho)$, the algorithm constructs for every $i\in [n]$ the set of cuboids $\C_i$, drops the first two coordinates that are fixed, and sets $\B_i$ to be the set of obtained intervals.
(In fact, this set is always a single interval in view of \cref{lem:removePeriod}.)
For the group $\sub$ we use \cref{lem:rangecoveraperRuntime} or \cref{lem:rangecoverperRuntime} depending on periodicity, and store the results as the sets $\S_i$.
The dimensions of the $2\CNF$ data structure constructed from $\Pairs$ are
\[
d_\ell=\begin{cases}
    1&\text{if }\bor=\{U\}\\
    1&\text{if }\bor=\bor(\rho)
\end{cases}
\text{ and }
d_r=\begin{cases}
    2&\text{if }\sub \text{ is not highly periodic }\\
    3&\text{if }\sub \text{ is highly periodic }
\end{cases}.
\]

\para{Query.} 
For a query $(S[1..b],S[x..y])$, the algorithm verifies that $U=S[1..b]$ is a border by checking if $\LCP_S^R(b,n)=b$.
If this is not the case, the oracle answers False.
Otherwise, the algorithm checks if $U$ is highly periodic by querying $\IPM_S$.
If it is not, the algorithm sets $\bor=\{U\}$ and checks if $f_\bor = \infty$.
If $f_\bor = \infty$, the algorithm returns True. 
Otherwise, it creates a $1$-dimensional point $p_\ell = (0)$.

If $U$ is highly $\rho$-periodic, the algorithm defines $\bor=\bor(\rho)$, and checks if $f_\bor = \infty$ or $q_\bor > |U|$. If this is the case, the algorithm returns True.
Otherwise, it sets $|U| = q\rho + d$ with the unique integers and $d \in [0.. \rho-1]$ and $q$.
The algorithm then creates the point $(0,d,q)$ and trims it to $p_\ell = (q)$ as the first two coordinates are fixed.

Now the algorithm works with $f=f_\bor$ (which is finite).
It queries $\IPM_S$ to check if there is an occurrence of $V = S[x..y]$ that covers $f$. 
Technically, the algorithm queries $\IPM_S(S[f-|V| +1 .. f+|V|-1],V)$ and checks if the output is empty.
If there is no such occurrence, the algorithm returns False.
Otherwise, the algorithm picks an arbitrary occurrence $i_V$ of $V$ covering $f$ and computes the integers $\ell$ and $r$ such that $V = S[f-\ell .. f+r]$.

Next, the algorithm checks if $V$ is highly periodic by querying $\IPM_S$.
Getting the result, it chooses the appropriate group $\sub$ for $V$ and creates a point for $V$.
If $V$ is not highly periodic, the point is $p_r = (\ell,r)$. If $V$ is highly $\rho$-periodic then the algorithm defines $\root$ and computes the point $p_r=(d_\ell,d_r,q_{\ell,r})$ for $V$  such that $V = \root[\rho - d_\ell + 1.. \rho]\cdot \root^{q_{\ell,r}} \cdot \root[1..d_r]$.
Finally, the algorithm queries the $2\CNF_{(\bor,\sub)}$ data structure with the point $(p_\ell,p_r)$ and returns the answer obtained for this query.

\para{Complexity.}
As in the prefix-suffix case, the construction time is dominated by the amount needed to build $2\CNF$ structure (all the rest is within $O(n\log^2n)$ bound).
As this structure is at most 4-dimensional, we get 
the $O(n\log^3n)$ bound from \cref{lem:2CNFDS}.
Since the number of such structures the algorithm builds for different pairs of groups is $O(\log^2n)$,  the total construction time is $O(n\log^5 n)$.

A query consists of a constant number of $\LCP$ queries, $\IPM$ queries, and arithmetic operations.
The algorithm also applies a single query to a $2\CNF$ data structure, which takes $O(\log^3n)$ time due to \cref{lem:2CNFDS}.

\para{Correctness.}
We claim that the query returns the correct answer to the question ``Is $(S[1..b], S[x..y])$ a border-substring cover?''.
If the algorithm returns False due to $S[1..b]$ not being a border, this is obviously correct.
Assume that $U=S[1..b]$ is a border and let $\bor$ be its group.
If the query returns True due to $f_\bor=\infty$ or 
Let $|U|<q_\bor$, then $U$ is a 1-cover, and then forms a 2-cover with any substring.

In the remaining case the query returns the same answer as the $2\CNF$ data structure built for the groups of $U$ and $S[x..y]$. 
Here the analysis is similar to the prefix-suffix case, so we omit it.

\section{Free Points Data Structure}\label{app:FreePoint}

In this section, we prove \cref{lem:AllFreePoints}.
In the proof of \cref{lem:AllFreePoints} we use the concept of \emph{persistent data structure} \cite{DSST89}.
A data structure is persistent if it supports multiple versions of itself and allows quick access to any version for querying, deletion, or update (an update creates a new version).

We first recall the definition of a free point and the statement of \cref{lem:AllFreePoints}.
\deffreepoint*

\lemAllFreePoints*
\begin{proof}
    The main point of the algorithm is an efficient reduction of the 2-dimensional problem to its 1-dimensional analog.
    The algorithm uses an auxiliary tree $\cT$ and the main structure $\cD$, which is a variant of persistent lazy segment tree \cite{RuSh17}.
    The details are described below.

    We take a fully balanced binary tree with $2^{\ceil{\log n}}$ leaves and delete $2^{\ceil{\log n}}-n$ rightmost leaves together with all internal nodes having no leaves remained in their subtrees.
    The remaining leaves are enumerated from $1$ to $n$ in a natural order.
    Thus, the leaves in a subtree of every node form a sub-range of $[n]$; we use these ranges as names of nodes.
    The only data stored in a node is the links $.\mathit{left}$ and $.\mathit{right}$ to its children.
    We refer to the obtained tree $\cT_0$ as the \emph{blank} tree.
    It is used to construct both $\cT$ and $\cD$.

    Building $\cT$ from $\cT_0$, we interpret each leaf as the $y$-coordinate of a point in $[n]^2$; the leaf $k$ corresponds to all points $(x,y)\in[n]^2$ with $y=k$.
    The algorithm processes each rectangle $R\in \R$, adding the information about its $x$-range to the nodes of $\cT_0$.
    Precisely, a rectangle $[x_1..x_2]\times[y_1..y_2]$ is fed to the following recursive procedure, starting at the root $[n]$ of $\cT_0$:
    \begin{itemize}
        \item for the current node $[y'..y'']$:
        \begin{itemize}
            \item if $[y'..y'']\cap [y_1..y_2]=\varnothing$, stop;
            \item if $[y'..y'']\subseteq [y_1..y_2]$, add the range $[x_1..x_2]$ to $[y'..y'']$ and stop;
            \item otherwise, call the procedure for both children of $[y'..y'']$.
        \end{itemize}
    \end{itemize}
    A well-known observation says that the procedure is called for at most $4$ nodes at each level of $\cT_0$.
    Therefore, each rectangle is processed in $O(\log n)$ time and adds $O(\log n)$ to the space used by $\cT_0$.
    After processing all rectangles from $\R$ in $O(n\log n)$ time, the resulting tree is $\cT$.
    This tree stores stores $O(n\log n)$ information.

    \begin{claim}\label{clm:xranges}
        A point $(x,y)$ is $\R$-free if and only if $x$ does not belong to any $x$-range stored in a node on the path from the root of $\cT$ to the leaf $y$.
    \end{claim}
    \begin{claimproof}
        Note that the algorithm partitions each rectangle $R\in \R$ into a set of $O(\log n)$ disjoint ``new'' rectangles with the same $x$-range such that the $y$-ranges of these rectangles correspond to some nodes of $\cT$.
        Then a point is $\R$-free if and only if it is not contained in any of the new rectangles. 
        Each node $[y_1..y_2]$ stores all new rectangles of the form $[x_1..x_2]\times[y_1..y_2]$.
        Hence, the nodes on the path from the root to $y$ store all new rectangles containing $y$ in their $y$-range.
        Then $(x,y)$ is $\R$-free if and only if $x$ is out of all $x$-ranges stored in these nodes.
    \end{claimproof}
    By \cref{clm:xranges}, to report all free points it suffices to find, for each $y\in[n]$, the complement of the union of all $x$-ranges stored in the nodes on the path from the root of $\cT$ to the leaf $y$.
    First we describe how to compute the complement of the union of a fixed set $\cX$ of $x$-ranges.

    \para{Solution for the $1$-dimensional case.} The algorithm starts with $\cT_0$, associates each leaf with the $x$-coordinate of a point, and processes ranges from $\cX$ one by one.
    Each node $[x_1..x_2]$ stores a single value $[x_1..x_2].f$, which is the number of points from $[x_1..x_2]$ that are not contained in already processed ranges.
    When all ranges are processed, points in $\overline{X}=[n]\setminus \bigcup_{X\in\cX} X$ are reported. 
    We assume that the algorithm reports \emph{all} points, and consider the other two options stated in the lemma in the end of the proof.
    Reporting is made during a partial depth-first traversal of the obtained tree $\cT_\cX$.
    This traversal skips subtrees rooted at zero-valued nodes and thus visits only $O(|\overline X|\log n)$ nodes, including $|\overline X|$ leaves. 
    Hence the reporting costs $O(|\overline X|\log n)=O(\Output\cdot\log n)$ time.
    Let us describe the processing phase.

    The value of a node is initialized as the number of points in its range.
    The call $\Add([n],X)$ is then performed for each $X\in\cX$, where the recursive function $\Add$ is defined as follows.
    \begin{itemize}
        \item $\Add(\mathsf{node}\ [x_1..x_2], \mathsf{range}\ X):$
        \begin{itemize}
            \item If $[x_1..x_2]\subseteq X$, set $[x_1..x_2].f=0$
            \item If $[x_1..x_2]\cap X\ne \varnothing$ and $[x_1..x_2].f>0$, set $[x_1..x_2].f= \Add([x_1..x_2].\mathit{left}) + \Add([x_1..x_2].\mathit{right})$
            \item Return $[x_1..x_2].f$
        \end{itemize}
    \end{itemize}
    If a value $[x_1..x_2].f$ equals $|[n]\setminus (X_1\cup\cdots\cup X_k)|$ after processing the ranges $X_1,\ldots, X_k$, we call it \emph{correct}.
    The following claim proves correctness of values in all nodes traversed during the reporting phase.
    \begin{claim} \label{clm:1dimensional}
        After any sequence of calls $\Add([n],X_1),\ldots, \Add([n],X_k)$, every node $[x_1..x_2]\in \cT_\cX$  has either a correct value or a zero-valued proper ancestor.
    \end{claim}
    \begin{claimproof}
        The proof is by induction on $k$.
        The initialization of values proves the base case.
        For the step case, assume that the claim is true after processing $X_i$ and consider the call $\Add([n],X_{i+1})$.
        Since zero values never change and the recursion does not reach below a zero-valued node, it is sufficient to consider an arbitrary node $[x_1..x_2]$ with nonzero values in it and in all its ancestors.
        If $[x_1..x_2]\subseteq X_{i+1}$, then the value of either this node or one of its ancestors is correctly set to $0$ during the call.
        If $[x_1..x_2]\cap X_{i+1}=\varnothing$, the value $[x_1..x_2].f$ remains unchanged which is correct by the inductive hypothesis.
        In both cases, the step case holds.
        In the remaining case, the value $[x_1..x_2].f$ was set to the sum of the values of its children.
        If an incorrect value was assigned, then \emph{earlier} during the call another incorrect value was assigned to a child of $[x_1..x_2]$.
        But if an incorrect assignment is necessarily preceded by another incorrect assignment, then no incorrect assignment can happen.
        Hence, all values set as the sums of their children's values are assigned correctly. 
        The step case is proved.
    \end{claimproof}
    By \cref{clm:1dimensional}, in the reporting phase the algorithm meets only correct values and thus correctly reports the set $\overline{} X$.
    During the call $\Add([n],X)$, at most $4$ nodes at each level are touched.
    Thus, each range $X\in \cX$ is processed in $O(\log n)$ time, and the whole processing phase requires $O(|\cX|\log n)$ time.

\para{Processing of all $x$-ranges stored in $\cT$.}
    For a node $t\in\cT$, let $\cX_t$ be the set of $x$-ranges stored in all nodes on the path from the root to $t$.
    The algorithm traverses $\cT$ depth first, creating the tree $\cT_{\cX_t}$ on the first visit to $t$ and deleting it on the last visit.
    When $t=y$ is a leaf, the reporting phase is run for $\cT_{\cX_y}$, during which $\R$-free points with the second coordinate $y$ are reported.
    In order to save space and construction time, the trees $\cT_{\cX_t}$ are built and stored as versions of a persistent data structure $\cD$.

    Let us briefly explain how it works.
    We first build the tree $\cT_{\cX_{[n]}}$, processing all ranges stored in the root of $\cT$.
    To every node of this tree we assign its \emph{version index}, equal to 0; version 0 is now the \emph{current version}.
    Then we start a depth-first traversal of $\cT$, adding and deleting versions as follows.
    
    The index of the current version is always the depth of the \emph{currently traversed node} of $\cT$.
    To \emph{copy} a node $v$ means to create a new node $v'$ with $v'.\mathit{left}=v.\mathit{left}$, $v'.\mathit{right}=v.\mathit{right}$, $v'.f=v.f$, and $v'.\mathit{version}$ being the current version.
    Reaching node $t'$ from its parent $t$, we create a new version as follows.
    We copy the root $v$ of the version $\cT_{\cX_t}$ into $v'$.
    Now we can navigate $\cT_{\cX_t}$ using the root $v'$ instead of $v$.
    Starting from $v'$, we process all ranges stored in $t'$ with with a simple modification of the function $\Add(\cdot,\cdot)$.
    This modification, when calling to a child $x$ of the currently processed node $u$, checks if $x.\mathit{version}$ is current; if not, it copies $x$ to $x'$, updates the link of $u$ from $x$ to $x'$, and then runs the recursive call for $x'$. 
    When all ranges stored in $t'$ are processed, we have the tree $\cT_{\cX_{t'}}$ with the root $v'$.

    When visiting $t$ for the last time, the algorithm deletes all nodes of the current version, starting from the root.
    The time spent for all deletions in the same as for all creations of new copies,
    so we can ignore it.
    Note that at every moment all existing versions correspond to the nodes on the path from the root to the current node of $\cT$.
    In particular, all version indices are different.
    In order to return to the previous version after deleting the current one, it suffices to store all roots in an array of version indices.
    
    Every call to the modified function $\Add([n],\cdot)$ still takes $O(\log n)$ time as in the original implementation, since what we added is just $O(1)$ copying operations per level of the tree (recall that at most 4 nodes per level of the tree are touched during this call).
    As $\cT$ contains $O(n\log n)$ ranges, and each range is added to $\cD$ only once, the processing time is $O(n\log^2 n)$.
    As every two reporting phases output disjoint sets of points, the reporting costs $O(n+\Output\cdot\log n)$ time.
    This gives the time bounds stated in the lemma.

    It remains to consider the other two modes of reporting: report the minimal $x$ for each $y$ and report all pairs with $(x+y)\le m$.
    As we traverse the leaves of any tree $\cT_\cX$ in increasing order, we just stop the depth-first traversal of the reporting phase at the moment when all required pairs $(x,y)$ with the given $y$ are reported.
    Thus, the reporting costs the same $O(n+\Output\cdot\log n)$ time as in the general case.
    If we report at most one $x$ for each $y$, then $\Output=O(n)$.
    The lemma is proved.
\end{proof}

\section{Reporting All 2-Covers Up To A Given Length}\label{app:report}

In this section, we prove \cref{thm:reporting}.
The algorithm operates in two phases. 
First, the algorithm reports all non-highly periodic $2$-covers (see \cref{sec:reportaper}).
Then, the algorithm uses the non-highly periodic $2$-covers to find all the highly periodic covers (see \cref{sec:reportper}).

\subsection{Report All Non-Highly Periodic 2-Covers}\label{sec:reportaper}
In this section we prove the following lemma.

\begin{lemma}\label{lem:reportingAperiod}
    Let $S$ be a string.
    There exists an algorithm that reports all non-highly periodic $2$-covers of $S$ (and may report also some highly periodic $2$-covers as well) in $O(n\log^{4}n+\Output\cdot \log n)$ time.
\end{lemma}

We split the proof of \cref{lem:reportingAperiod} into two parts, one for reporting prefix-suffix $2$-covers, and one for reporting border-substring $2$-covers.

\subsubsection{Report All Non-Highly Periodic Prefix-Suffix 2-Covers}
The following lemma is useful in this section.
In essence, we claim that the dimension of the ranges of \cref{lem:rangecoveraperRuntime} are actually $1$ (and not $2$) in the case in which $f$ is an endpoint of $S$.
\begin{lemma}\label{lem:rangeReduction}
    Let $i\in[n]$ be an index and let $k\in\N$.
    There exists a set $\I_\ell$  of $O(1)$ intervals such that for any $z\in\N$ where  $\pref=S[1..1+z]$ has $z\in[1.5^k..1.5^{k+1}]$ we have:
    \begin{enumerate}
        \item If $z \in\bigcup_{I\in\I_\ell }I$, then $\pref$ covers $i$.
        \item If $\pref$ is a non-highly periodic string covering $i$, then $z \in\bigcup_{I\in\I_\ell}I$.
    \end{enumerate}

    Symmetrically, there exists a set $\I_r$ of $O(1)$ intervals such that for any $z\in\N$ where  $\suff= S[n-z .. n]$ has $z\in[1.5^k..1.5^{k+1}]$ we have:
    \begin{enumerate}
        \item If $z \in\bigcup_{I\in\I_r }I$, then $\suff$ covers $i$.
        \item If $\suff$ is a non-highly periodic string covering $i$, then $z \in\bigcup_{I\in\I_r}I$.
    \end{enumerate}
    
    Moreover, $\I_\ell$ and $\I_r$ can be computed in $O(\log^2 n)$ time.
\end{lemma}
\begin{proof}
    We use \cref{lem:rangecoveraperRuntime} with $f=1$, $i$ and $k$ to obtain a set $\R_\ell$ of rectangles.
    Let $\I_\ell=\{[a..b]\mid [c..d]\times[a..b]\in\R \textit{ and }0\in [c..d] \}$.
    We use again \cref{lem:rangecoveraperRuntime} with $f=n$, $i$ and $k$ to obtain a set $\R_r$ of rectangles.
    Let $\I_r=\{[a..b]\mid \exists_c[a..b]\times[c..d]\in\R \textit{ and }0\in [c..d] \}$.

    Note that every point $(\ell,r)$ such that $S[1-\ell..1+r]=S[1..1+z]$ must have $\ell=0$ and $z=r$.
    Therefore, $z\in\bigcup_{I\in\I }I$ if and only if $(0,z)\in\bigcup_{R\in\R_r }R$. 
    The claim immediately follows from \cref{lem:rangecoveraperRuntime} (the proof for $\I_r$ is symmetric).
\end{proof}

Now, we are ready to prove the following lemma, which yields the reporting mechanism stated in \cref{lem:reportingAperiod} for \emph{prefix-suffix} $2$-covers.
\begin{lemma}\label{lem:reportingAperiodPS}
    Let $S$ be a string and $m$ an integer.
    There is an algorithm that reports a set $\C'_m$ of $2$-covers of length at most $m$ such that every non non-highly periodic prefix-suffix $2$-covers of $S$ is in $\C'_m$ in $O(n\log^{4}n+\Output\cdot \log n)$ time.
\end{lemma}
\begin{proof}
The algorithm starts with the common preprocess phase described in \cref{sec:ranges}.
The algorithm iterates over all pairs of integers $(k_\ell,k_r)$ such that $1.5^{k_\ell}\le n$ and $1.5^{k_r}\le n$.
Let $(k_\ell,k_r)$ be such a pair, the algorithm processes $(k_\ell,k_r)$ as follows.
For every $i\in[n]$
the algorithm applies \cref{lem:rangeReduction} with $k=k_\ell$ to get $\I_{\ell}$ and with $k=k_r$ to get $\I_r$.
Let $\R_i=\{I\times [-\infty..\infty]\mid I\in\I_\ell\}\cup \{[-\infty..\infty]\times I\mid I\in\I_r\}$.
The algorithm uses \cref{lem:inverseofhypercubes} to compute a set $\overline{\R}_i$ of rectangles such that   $\bigcup_{R\in \overline{\R}_i}R = [-\infty .. \infty]^2 \setminus \bigcup_{R \in \R_i}R$ in $O(1)$ time.
Let $\Valid=\{[1.5^{k_\ell}..1.5^{k_\ell+1}]\times [1.5^{k_r}..1.5^{k_r+1}]\}$, and let $\overline{\Valid}$ be the inverse set of rectangles computed by \cref{lem:inverseofhypercubes}. 
Then, the algorithm computes $\overline{\R}=(\bigcup_{i\in[n]}\overline{\R_i})\cup (\overline{\Valid})$.
Finally, the algorithm uses the third component of \cref{lem:AllFreePoints} with $\overline{\R}$ and $m-2$ as a bound for $x+y$.
For every free point $(\ell,r)$ with respect to $\overline{\R}$ reported by \cref{lem:AllFreePoints}, the algorithm reports $(S[1..1+\ell],S[n-r..n])$ as a $2$-cover of $S$.

\para{Correctness.}
Let $(\pref,\suff)$ be a non-highly periodic $2$-cover with $\pref = S[1..1+\ell]$, $\suff = S[n-r..n]$ and $\ell+r+2 \le m$.
Let $k_\ell$ and $k_r$ be the integers such that $\ell \in [1.5^{k_\ell} .. 1.5^{k_\ell+1}]$ and $r\in[1.5^{k_r}..1.5^{k_r+1}]$.
Clearly, $(\ell,r)\in [1.5^{k_\ell}..1.5^{k_\ell+1}]\times [1.5^{k_r}..1.5^{k_r+1}]$.
Additionally, since $(\pref,\suff)$ covers $i$ for every $i\in [n]$, it must be the case that $(\ell,r) \notin R$ for every $R \in \overline{\R_i}$ by \cref{lem:rangeReduction}.
It follows that $(\ell,r)$ is a free point with respect to $\overline{\R}$ with $\ell+r \le m-2$, as required.

Let $(\pref,\suff)$ be a pair reported by the algorithm with  $\pref = S[1..1+\ell]$ and $\suff = S[n-r..n]$.
Let $(k_\ell,k_r)$ be the pair such that $(\pref,\suff)$ was reported by the instance of \cref{lem:AllFreePoints} created when processing $(k_\ell,k_r)$.
Let $\overline{\R}$ and $\overline{\Valid}$ be the sets computed when processing $(k_\ell,k_r)$.
Since $(\pref,\suff)$ was reported, it must be the case that $(\ell,r)$ is free with respect to $\overline{\R}$.
Since $\overline{\Valid}\subseteq \overline{\R}$ it must be that $(\ell,r)\in\Valid$ which means $\ell \in [1.5^{k_\ell} .. 1.5^{k_\ell+1}]$ and $r\in[1.5^{k_r}..1.5^{k_r+1}]$.
Moreover, for every $i\in[n]$ we have $(\ell,r) \notin \bigcup_{R \in \overline{\R_i}}R$ and therefore $(\ell,r)\in\bigcup_{R \in \R_i}R$.
It follows from \cref{lem:rangeReduction} that $(\pref,\suff)$ covers $i$.
It follows that $(\pref,\suff)$ is a $2$-cover.
Finally, the point $(\ell,r)$ satisfies $\ell + r \le m-2$ which leads to $|\pref| + |\suff| = \ell+r+2 \le m$ as required.

\para{Complexity.}
The preprocess of \cref{sec:ranges} is computed in $O(n\log^2n)$ time.
When a pair $(k_\ell,k_r)$ is processed, \cref{lem:rangeReduction} and \cref{lem:inverseofhypercubes} are applied $O(n)$ times, which takes $O(n\log^2n)$ time.
Then, \cref{lem:AllFreePoints} is applied on a set of $\Theta(n)$ rectangles, which takes $O(n\log^3n + \Output_{(k_\ell,k_r)} \cdot \log^2 n)$ time, with $\Output_{(k_\ell,k_r)}$ being the set of free points with respect to the rectangles $\overline{\R}$ created for the pair $(k_\ell,k_r)$.
Due to the inclusion of $\overline{\Valid}$ in $\overline{\R}$, we have that every reported point $(\ell,r)$ has $\ell \in [1.5^{k_\ell}..1.5^{k_\ell+1}]$ and $r\in [1.5^{k_r}..1.5^{k_r+1}]$.
It immediately follows that every point is reported at most once.
It follows from \cref{lem:rangeReduction} that every free point $(\ell,r)$ with respect to $\overline{\R}$ corresponds to a $2$-cover $(S[1..1+\ell],S[n-r..n])$.
Notice that the same $2$-cover may be reported at most twice - in the case in which $S[1..1+\ell] = S[n-\ell .. n]$ and $S[1..1+r] = S[n-r..n]$.
It follows that the accumulated size of $\Output_{(k_\ell,k_r)}$ across all pairs $(k_\ell,k_r)$ is bounded by $2\cdot \Output$, with $\Output$ being the set of $2$-covers reported by the algorithm.
In conclusion, the total running time is bounded by $O(n\log^4 + \Output \cdot \log n)$ due to the existence of $O(\log^2n)$ pairs $(k_\ell,k_r)$.
\end{proof}

\subsubsection{Report All Non-Highly Periodic Border-Substring 2-Covers}\label{sec:reportall}

We proceed to prove that all non-highly periodic border-substring $2$-covers can be reported efficiently.
\begin{lemma}\label{lem:reportingAperiodBS}
    Let $S$ be a string and $m$ be an integer.
    There exists an algorithm that reports a set $\C_m$ of $2$-covers with length at most $m$, such that every non-highly periodic border-substring $2$-covers of $S$ is in $\C_m$.
    The running time of the algorithm is $O(n\log^4n + \Output \cdot \log^2 n)$.
\end{lemma}
\begin{proof}
    The algorithm starts with the common preprocessing phase described in \cref{sec:ranges}.
    We also assume that all $1$-covers are found in advance via the linear time algorithm of Moore and Smyth~\cite{MS94,MS95}..
    Every $2$-cover containing a $1$-cover is implicitly reported via the corresponding $1$-cover.
    The algorithm iterates every pair $(b,k)$ such that $S[1..b]$ is a non-highly periodic border and $k$ is an integer such that $1.5^k \le n$ as follows.
    The algorithm finds all occurrences of $S[1..b]$ in $S$ using \cref{lem:PM}.
    If every index in $S$ is covered by $S[1..b]$, the algorithm simply ignores $S[1..b]$, as all pairs including it are implicitly reported as the $1$-cover $S[1..b]$
    If $S[1..b]$ is not a $1$-cover, the algorithm picks an arbitrary index $f$ not covered by $S[1..b]$.
    For every index $i$ that is not covered by $S[1..b]$, the algorithm applies \cref{lem:rangecoveraperRuntime} with $f$,$i$, and $k$ to obtain a set $\R_i$ of rectangles.
    The algorithm then applies \cref{lem:inverseofhypercubes} to obtain a set of $O(1)$ rectangles $\overline{\R}_i$ such that $\bigcup_{R\in \overline{\R}_i}R = [-\infty .. \infty]^2 \setminus \bigcup_{R\in \R_i}R$,
    The algorithm then creates a set $\overline{\Valid}$ of $O(n)$ rectangles such that a point $(\ell,r)$ is not in a rectangle of $\overline{\Valid}$ if and only if $\ell+r+1 \in [1.5^k .. 1.5^{k+1}]$.
    Note that there is a set of $O(1.5^k) = O(n)$ rectangles that satisfy this constraint.
    Finally, the algorithm applies the third variant of \cref{lem:AllFreePoints} on the set $\overline{R} =  (\bigcup_{i \in U}\overline{\R}_i) \cup (\overline{\Valid})$ with $U$ being the set of indices not covered by $S[1..b]$ and $m-1 - b$ the bound on $x+y$.
    For every free point $(\ell,r)$ obtained by \cref{lem:AllFreePoints}, the algorithm reports the pair $(S[1..b], S[f-\ell ..f+r])$ as a $2$-cover.
\para{Correctness}
Let $(\bor,\sub)$ be a non-highly periodic border-substring $2$-cover with $\bor = S[1..b]$ and $b+|\sub| \le m$.
If every index in $S$ is covered by an occurrence of $\bor$, the pair is report implicitly via the $1$-cover $\bor$.
Otherwise, the index $f$ not covered by $\bor$ picked by the algorithm must be covered by $\sub$.
It follows that $\sub = [f-\ell .. f+r]$ for some $\ell,r\in\N$.
   Let $k$ be the integer such that $|\sub| = \ell+r-1 \in [1.5^k .. 1.5^{k+1}]$.
    Let $\overline{\R}$ be the set of rectangles on which the algorithm applied \cref{lem:AllFreePoints} when the pair $(b,k)$ was processed.
    It is clear from the definition of $\overline{\Valid}$ that the point $(\ell,r)$ is not in a rectangle of $\overline{\Valid}$.
    Note that every index in $S$ not covered by  $\bor$ must be covered by $\sub$.
    It follows from \cref{lem:rangecoveraperRuntime} and from $|\sub| \in [1.5^k..1.5^{k+1}]$ that $(\ell,r)$ is a free point with respect to $\overline{\R}$, and therefore the pair $(S[1..b],S[f-\ell..f+r])$ is reported as a $2$-cover.
    Finally the point $(\ell+r)$ respects the bound of the third variant of \cref{def:freepoint}, since $|\bor| + |\sub| = b + r + \ell + 1 \le m$ as required.

    Now, let $(S[1..b],S[f-\ell .. f+r])$ be a pair reported by the algorithm.
    Let $k$ be the unique integer such that $\ell+r-1 \in [1.5^k .. 1.5^{k+1}]$.
    Since the pair $(S[1..b],S[f-\ell..f+r]$ was reported by the algorithm, it must be the case that $(\ell,r)$ is a free point with respect to a set of rectangles $\overline{\R}$ created when processing some pair $(b,k')$.
    Due to the inclusion of $\overline{\Valid}$ in $\overline{\R}$ and the uniqueness of $k$, it must be the case that $k'=k$.
    According to \cref{lem:rangecoveraperRuntime}, it follows from $(\ell,r)$ being a free point with respect to $\overline{\R}$ and from $\ell+r+1 \in [1.5^k .. 1.5^{k+1}]$ that $S[f-\ell..f+r]$ covers every index that is not covered by $S[1..b]$, so $(S[1..b],S[f-\ell..f+r])$ is indeed a $2$-cover.
    Finally, the bound on $\ell + r$ on any reported point ensures that $|S[1..b]| + S[f-\ell .. f+r]| = b+\ell+r+1 \le m$ as required.
\para{Complexity}
The preprocessing of \cref{sec:ranges} is carried in $O(n\log^2n)$ time.
The set of all non-highly periodic borders is computed in $O(n)$ time as in \cref{sec:bsoracle}.
For every $1$-cover $S[1..b]$ identified the algorithm, the algorithm applies \cref{lem:allsubstrings} to report all $2$-covers containing $S[1..b]$ in time $O(\Output_b)$ with $\Output_b$ being the number of such borders. 
We proceed to consider non-highly periodic borders $S[1..b]$ that are not $1$-covers.
For every pair $(b,k)$, the algorithm applies pattern matching once, and creates $O(n)$ rectangles using \cref{lem:rangecoveraperRuntime}.
The time complexity of these operations sums up to $O(n\log^2n)$.
Then, the algorithm finds free point on a set of $\Theta(n)$ rectangles using \cref{lem:AllFreePoints}.
The time complexity of this part of the algorithm is $O(n\log^3n + \log^2 n \cdot \Output_{b,k})$ with $\Output_{b,k}$ being the number of free points with respect to the rectangle set $\overline{\R}$ created when processing $(b,k)$.
We bound the number of times that a $2$-cover $(S[1..b],\sub)$ can be reported by the algorithm via a free point.
Let $k$ be the unique integer such that $|\sub| \in [1.5^k..1.5^{k+1}]$.
When the pair $(b,k)$ is being processed, $(S[1..b],\sub)$ may be reported multiple times due to multiple occurrences of $\sub$ in the proximity of $f$.
More specifically, there may be two (or more) points $(\ell_1,r_1)$ and $(\ell_2,r_2)$ such that $S[f-\ell_1 ..f+r_1] = S[f-\ell_2 ..f+ r_2]=\sub$.
We claim that there are at most $6$ such points. This is since due to \cref{lem:rangecoveraperRuntime}, we have $\per(\sub) \ge \frac{1.5^k}{4}$ (due to $(\ell,r)$ being a free point with respect to $\overline{\R}$).
It follows that two starting indices of occurrences of $\sub$ must be at least $\frac{1.5^k}{4}$ indices apart.
Since $|\sub| \le 1.5^{k+1}$ we have that at most $1.5^{k+1} / \frac{1.5^k}{4} = 6$ occurrences of $\sub$ can touch the index $f$.
The $2$-cover $(S[1..b],\sub)$ can also be reported with reversed roles i.e. with $S[1..b]$ acting as the substring and $\sub$ as the border.
This can only happen when processing the pair $(b',k')$ with $k'$ being the unique integer such that $b \in [1.5^{k'} .. 1.5^{k'+1}]$ and $S[1..b'] = \sub$.
Due to the same reasoning as before, the $2$-cover can be reported at most $O(1)$ times when the pair $(b',k')$ is processed.
It follows from the above analysis that across all pairs $(b,k)$, a $2$-cover can be reported at most $12 = O(1)$ times.
It follows that the total contribution of the $O(\log^2 n \cdot  \Output_{(b,k)})$ component to the time complexity is $O(\log^2 n \cdot \Output )$.
In conclusion, the algorithm runs in time $O(n \log^{4}n + \log n \cdot \Output)$ (due to the existence of $O(\log^2 n)$ pairs $(b,k)$).
\end{proof}

Note that \cref{lem:reportingAperiod} follows directly from \cref{lem:reportingAperiodPS,lem:reportingAperiodBS}.

\subsection{Report All Highly-Periodic 2-Covers}\label{sec:reportper}
We are now ready to prove \cref{thm:reporting}.
\begin{proof}[Proof of \cref{thm:reporting}]
    The algorithm starts by applying \cref{lem:reportingAperiod} to obtain and report every non-highly periodic $2$-cover of $S$ with length bounded by $m$.
    In addition, the algorithm constructs the $2$-cover oracle of \cref{thm:oracle}.
    In particular, we assume that the algorithm has access to $\Dict$ and $\Dict_q$ presented in \cref{sec:bsoracle}. 
    The key idea for obtaining the rest of the $2$-covers is attempting to extend short periodic strings components of non-highly periodic $2$-covers.

    \para{Prefix-Suffix.}
    To streamline the presentation, we present two operators for extending a periodic substring.
    \begin{definition}
        Let $S$ be a string and let $\pref = S[1..p]$ be a $\rho$-periodic prefix of $S$.
        The Operator $\PrefExtend(\pref)$ is defined as follows.
        \[\PrefExtend(\pref)=\begin{cases}
    S[1..p+\rho] & \textsf{if }  S[1..p+\rho] \textsf{ is } \rho \textsf{-periodic}\\
    \EndSymbol &\text{otherwise.}
\end{cases}\]
     Similarly, the operator $\SufffExtend(.)$ is defined on a $\rho$-periodic suffix $\suff=S[s..n]$ as follows.
    \[\SufffExtend(\pref)=\begin{cases}
    S[s-\rho..n] & \textsf{if } S[s-\rho .. n] \textsf{ is } \rho \textsf{-periodic}\\
    \EndSymbol &\text{otherwise.}
    \end{cases}\]
    \end{definition}
     We slightly abuse notation by interpreting $\PrefExtend^0(X)$ (i.e applying the operator zero times on $X$) as $X$ even for an aperiodic prefix $X$, on which the operator is not defined (a similar abuse is allowed for suffixes).
    Observe that an $\IPM$ data structure of \cref{lem:IPM} can be used to obtain $\PrefExtend$ and $\SufffExtend$ in constant time.

    Finally, we make the following observation which is a direct implication of \cref{lem:removePeriod}.
    \begin{observation}\label{obs:extendprefixsuffix}
        Let $(X,Y)$ be a highly periodic prefix-suffix $2$-cover of $S$.
        There is a unique core prefix-suffix $2$-cover $(X',Y')$ and two non-negative integers $p,s$ such that $\PrefExtend^p(X') =X$ and $\SufffExtend^s(Y') = Y$. 
        
        Furthermore, for every $(s',p') \in [0.. s] \times [0..p]$ the pair $(\PrefExtend^{p'}(X'),\SufffExtend^{s'}(Y'))$ is also a $2$-cover. 
    \end{observation}

    The uniqueness of $(X',Y')$ arises from the fact that both operations preserve the period of the prefix/suffix when applied to a periodic prefix/suffix.
    Therefore, in order to obtain $X$ from some $X'$, we either have to start from $X'=X$ and apply $\PrefExtend$ zero times if $X$ is aperiodic, or start with a short periodic prefix $X'$ with the same period as $X$ (which is unique). 
    The same reasoning applies to $Y$.
    
    The algorithm processes every non-highly periodic prefix-suffix $2$-cover $(\pref = S[1..p] ,\suff = S[n-s+1..n])$ with the following goal:
    Report all prefix suffix $2$-covers $(X,Y)$ with length bounded by $m$ such that $X= \PrefExtend^p(\pref)$ and $Y = \SufffExtend^s(\suff)$ for some naturals $p$ and $s$.
    It follows directly from \cref{obs:extendprefixsuffix} that all required prefix-suffix two covers are found by applying such process to every core pair.
    
    We proceed to describe the algorithm for processing a core pair $(\pref,\suff)$.
    \begin{itemize}
        \item     If both $\pref$ and $\suff$ are aperiodic - the algorithm does not apply further processing to it.
        \item    If $\pref$ is short $\rho$-periodic and $\suff$ is aperiodic, the algorithm initialize an iterator $p = 1$ initiates a loop. 
        In every step of the loop, the algorithm sets $X_p = \PrefExtend^p(\pref)$ and checks if 
         $X_p \neq \EndSymbol$, $|X_p|+|\suff| \le m$ and $(X_p,\suff)$ is a $2$-cover of $S$.
         If all of the conditions are satisfied, the algorithm reports $(X_p,\suff)$ as a $2$-cover, assigns $p\leftarrow p+1$ and repeats the loop.
         If one of the conditions is false - the loop is terminated.

    \item   If  $\pref$ is aperiodic and $\suff$ is short $\rho$-periodic  the algorithm processes $(\pref,\suff)$  in a symmetric manner to the previous case, extending $\suff$ using $\SufffExtend$.
    \item If both $\pref$ and $\suff$ are short periodic, the algorithm applies the extensions of the previous two cases in a nested loop fashion as follows.
    The algorithm initiates two iterators $p=0$ and $s=0$.
    In every step of the loop, the algorithm sets $X_p = \PrefExtend^p(X)$ and $Y_p = \SufffExtend^s(Y)$.
    The algorithm checks if $X_p \neq \EndSymbol$, $Y_p \neq \EndSymbol$, $|X_p| + |Y_p| \le m$ and $(X_p,Y_p)$ is a $2$-cover.
    If all conditions are satisfied - the algorithm reports $(X_p,Y_p)$ as a $2$-cover, assigns $p \leftarrow p+1$, and repeats the loop.
    If one of the conditions fails, and $p \neq 0$, the algorithm assigns $p \leftarrow 0$ and $s\leftarrow s+1$, and repeats the loop.
    If one of the conditions fails and $p=0$, the loop terminates.
    \end{itemize}

    \para{Correctness.}
    The correctness of the algorithm arises naturally from \cref{obs:extendprefixsuffix}.
    Clearly, every pair $(X,Y)$ reported by the algorithm is a $2$-cover with length bounded by $m$, as these conditions are explicitly tested for every reported pair.
    
    It remains to show that all required pairs are reported.
    Let $(X,Y)$ be a prefix suffix highly periodic $2$-cover with length bounded by $m$.
    Let $(X',Y')$ be the unique core pair derived from \cref{obs:extendprefixsuffix} and $p^*$ and $s^*$ the integers such that $X = \PrefExtend^{p^*}(X')$ and $Y=\SufffExtend^{s^*}(Y')$.
    From now on, we assume that both $X$ and $Y$ are highly periodic.
    The analysis required for the case in which only one of $X$ and $Y$ is highly periodic is derived from the analysis for the case in which both are highly periodic.
    In this case, both $X'$ and $Y'$ are short periodic with the same periods of $X$ and of $Y$, respectively.
    When the pair $(X',Y')$ is processed by the algorithm, we claim that for every $s' < s^*$ the pair $(X'_0,Y'_{s'})$ satisfies all conditions checked by the loop.
    \begin{enumerate}
        \item $X'_0 = X' \neq \EndSymbol$.
        \item $Y'_{s'} \neq \EndSymbol$ since $Y'_{s^*} \neq \EndSymbol$, and $\SufffExtend(\EndSymbol)$ is undefined. 
        \item $|X'_{0}| + |Y'_{s'}| \le m$. Since for every $s'<s^*$ such that $Y'_{s^*} \neq \EndSymbol$ and $Y'_{s'} \neq \EndSymbol$ we have $|Y'_{s'}| < |Y'_{s^*}|$, and we have $|X| + |Y| = |X'_{p^*}| + |Y'_{s^*}| \le m$.
        \item $(X'_0,Y'_{s'})$ is a $2$-cover due to \cref{obs:extendprefixsuffix}.
    \end{enumerate}
   It follows that the iterator $s$ of the loop will reach the value $s^*$.
   A similar analysis shows that when reaching $s^*$ and starting the process of increasing $p$, the $p$ iterator of the loop would reach the value $p^*$.
   At this point, the pair $(X,Y) = (X'_{p^*}, Y'_{s^*})$ is reported, as required.

\para{Complexity.}
We analyze the complexity of the nested loop executed in the case in which both $X$ and $Y$ are short periodic. 
The complexity of the rest of the cases arises from the same arguments.
When processing a pair $(X,Y)$, each step of every loop is dominated by the query to the oracle - Obtaining $\PrefExtend^p(X)$ or $\SufffExtend^s(Y)$ can be done in constant time using $\PrefExtend^{p-1}(X)$ of $\SufffExtend^{s-1}(Y)$ and the rest of the checks consist of integer comparisons.
Each $2$-oracle query on a pair $(X_p,Y_s)$ to the oracle that reports True can be charged on the reported pair $(X_p,Y_s)$.
Due to the uniqueness of the core pair $(X,Y)$ with respect to $(X_p,Y_s)$, the total number of such queries is $\Output$.
Each query on a pair $(X_p,Y_s)$ that reports False is charged on the pair $(X_0,Y_s)$.
If $p > 0$, it means that the pair $(X_0,Y_s)$ was reported as a $2$-cover.
It follows that the total number of such queries on pairs $(X_p,Y_s)$ with $p>0$ is bounded by $\Output$ across all core pairs.
If $p=0$, the loop terminates after the query.
It follows that there is at most one such query for every processed core pair, and their total number across all pairs is bounded by $\Output$ as well.
It follows that the overall complexity of processing all pairs is bounded by $O(\Output \cdot \log^3)$ as required.

\para{Border-Substring.}
    The algorithm proceeds to report all border-substring $2$-covers.
    The reporting is carried out following the same principles as in the prefix-suffix case.
    However, extending the periodic substrings requires a more careful treatment.

    We define the following operator for extending substrings.

     \begin{definition}
        Let $S$ be a string and let $\sub = S[i..j]$ be a $\rho$-periodic prefix of $S$.
        The Operator $\SubExtend(\sub)$ is defined as follows.
        \[\SubExtend(\sub)=\begin{cases}
    S[i..j+\rho] & \textsf{if }  S[i..j+\rho] \textsf{ is } \rho \textsf{-periodic}\\
     S[i-\rho..j] & \textsf{if }  S[i-\rho..j] \textsf{ is } \rho \textsf{-periodic and } S[i..j+\rho] \textsf{ is not} \\
    \EndSymbol &\text{otherwise.}
\end{cases}\]
    \end{definition}
    Again, we slightly abuse notation by allowing $\SubExtend^0(X) = X$, even for non-periodic substrings $X$.
    Note that unlike $\PrefExtend$ and $\SufffExtend$, the operator $\SubExtend$ is sensitive to the exact location of the substring $\sub$ in $S$.
    However, it is still easy to obtain $\SubExtend(\sub)$ from $\sub$ using $\IPM_S$ in constant time.

    The following observation parallels \cref{obs:extendprefixsuffix} and arises due to the same arguments.
    \begin{observation}\label{obs:extendbordersubstring}
        Let $(X,Y)$ be a highly periodic border-substring $2$-cover of $S$ such that $X$ is not a $1$-cover.
        Let $f$ be an index not covered by $X$.
        There is a unique core border-substring $2$-cover $(X',Y')$ and two non-negative integers $b,s$ such that $\PrefExtend^b(X') =X$ and $\SubExtend^s(Y') = Y$.
        Where the $\SubExtend$ operations applied to $Y'$ are with respect to an occurrence of $Y'$ covering $f$.
        
        Furthermore, for every $(s',p') \in [0.. s] \times [0..p]$ the pair $(\PrefExtend^{p'}(X'),\SufffExtend^{s'}(Y'))$ is also a $2$-cover. 
    \end{observation}

    Not that an occurrence of $Y'$ covering $f$ must exist, as $Y$ must cover $f$, and $Y'$ covers every index covered by $Y$ by \cref{lem:removePeriod}.

    As an initial step, the algorithm uses the $O(n)$ time algorithm of Smyth~\cite{S02} to find all $1$-covers of $S$.
    For every $1$-cover $\bor$ of $S$, all pairs $(\bor,\sub)$ are considered to be reported implicitly via $\bor$.
    
   We now proceed to treat the non-trivial case.
   As in the prefix-suffix , the algorithm processes every core pair $(X,Y)$ and reports all pairs that arise from $(X,Y)$ via \cref{obs:extendprefixsuffix}.
    \begin{itemize}
        \item 
    If both $\bor$ and $\sub$ are aperiodic, the algorithm does not apply any further processing.
    
   \item If $\bor$ is short $\rho$-periodic and $\sub$ is aperiodic, the algorithm attempts to extend $\bor$ similarly to the prefix extension process described in the prefix-suffix case with the following modification.
   As a preliminary step, the algorithm obtains $p^* = \Dict_q[b]$ (see \cref{sec:bsoracle}).
   Let $\bor = S[1..2\rho + d_b]$ for $d_b\in[0..\rho-1]$.
   The algorithm starts the process of extending $\bor$ from $\bor' = S[1..q^*\cdot \rho + d_b]$ i.e. setting the initial value of $p$ as $q^*$.
   In essence, the algorithm `skips' any extension of $\bor$ that is a $1$-cover.
   All $2$-covers containing a $1$-cover are reported implicitly.
   If $q^* = \infty$, the algorithm does not apply any further processing to the pair $(\bor,\sub)$.
   
    \item If $\sub$ is short $\rho$-periodic and $\bor$ is aperiodic, the algorithm attempts to extend $\sub$ as follows.
    Let $f=f_b$ be an index not covered by $S[1..b]$ extracted from $\Dict$ (see \cref{sec:bsoracle}).
    If $f=\infty$, $\bor$ is a $1$-cover and all pairs containing $\bor$ are reported implicitly.
    
    Otherwise, $f$ is an index not covered by $\bor$.
    The algorithm finds an occurrence of $\sub$ covering $f$ using the $\IPM_S$ data structure of \cref{lem:IPM}, let $\sub = S[f-\ell .. f+r]$ for some integers $\ell,r$ (note that such $\ell$ and $r$ must exist since $(\bor,\sub)$ is a $2$-cover).
    The algorithm starts the extension process from $Y_1 = S[f-\ell.. f+r]$.
    It initializes an iterator $b=1$, and the loop is carried as in the prefix suffix case, replacing $\PrefExtend$ with $\SubExtend$.
    
    \item If $\bor$ is short $\rho_b$-periodic and $\sub$ is short $\rho_s$-periodic, the algorithm processes $(\bor,\sub)$ in a nested loop fashion as follows.
    We present this case in more detail, to provide a full picture of the previous cases as well.
    First, the algorithm finds the minimal value $q$ such that $X_q = \PrefExtend^q(\bor)$ is not a $1$-cover. 
    This is done using $\Dict_q$.
    If no such value exists - the algorithm halts the processing of the pair $(\bor,\sub)$.
    If this value exists, the algorithm sets an iterator $b \leftarrow q$.
    Next, the algorithm uses $\Dict$ to find an index $f$ not covered by $X_q$ and applies internal pattern matching (\cref{lem:IPM}) on the text $S[f-|\sub| .. f+|\sub|]$ to find an occurrence of $\sub$.
    If no such occurrence is found, the algorithm terminates the process of the pair $(\bor,\sub)$.
    Otherwise, the algorithm uses the occurrence to find two non-negative integers $\ell,r$ such that $\sub = S[f-\ell .. f+r]$.
    The algorithm then sets $Y_0 = S[f-\ell ..f+r]$ and initializes a secondary iterator $s = 0$.
    Now, the algorithm starts running the loop.
    At every step of the loop, the algorithm sets $X_b = \PrefExtend^b(\bor)$ and $Y_s = \SubExtend^s(\sub)$.
    The algorithm then checks if $Y_s \neq \EndSymbol$, $X_b \neq \EndSymbol$, $X_b$ is a border, $|X_b|+ |Y_s| \le m$, and if $(X_b,Y_s)$ is a $2$-cover.
    If all the conditions are true, the algorithm reports the pair $(X_b,Y_s)$, assigns $s\leftarrow s+1$, and repeats the loop.
    If one of the conditions is False, and $s\neq 0$, the algorithm sets $s\leftarrow 0$ and $b \leftarrow b+1$ and repeats the loop.
    If one of the conditions is False and $s = 0$, the loop terminates. 
\end{itemize}
     
Correctness and complexity follow due to similar arguments as in the prefix-suffix case.

\para{Correctness.}
    The correctness of the algorithm arises naturally from \cref{obs:extendbordersubstring}.
    Clearly, every pair $(X_b,Y_s)$ reported by the algorithm is a $2$-cover with length bounded by $m$, as these conditions are explicitly tested for every reported pair.
    
    It remains to show that all required pairs are reported.
    Let $(X,Y)$ be a prefix suffix highly periodic $2$-cover with length bounded by $m$.
    If $X$ is a $1$-cover, the pair $(X,Y)$ reported implicitly via $X$.
    For the rest of the analysis we assume that $X$ is not a $1$-cover.
    Therefore, there is a unique $(X',Y')$ core pair derived from \cref{obs:extendbordersubstring} and $b^*$ and $s^*$ the integers such that $X = \PrefExtend^{b^*}(X')$ and $Y=\SufffExtend^{s^*}(Y')$.
    From now on, we assume that both $X$ and $Y$ are highly periodic.
    The analysis required for the case in which only one of $X$ and $Y$ is highly periodic is derived from the analysis for the case in which both are highly periodic.
    In this case, both $X'$ and $Y'$ are short periodic with the same periods of $X$ and of $Y$, respectively.
    When the pair $(X',Y')$ is processed by the algorithm the initial value of the iterator $b$ is set to $q$ the minimal integer such that $\PrefExtend^q(\bor)$ is not a $1$-cover.
    By definition, $q \le b^*$.
    The algorithm then retrieves an index $f$ not covered by $X_q$ from $\Dict$.
    Note that $f$ is also not covered by $X$ due to \cref{lem:removePeriod}.
    There must be an occurrence of $\sub$ covering the index $f$, since there is an occurrence of $Y$ covering $f$, and $\sub$ is obtainable from $Y$ by removing $\per(Y)$ a certain number of times (\cref{lem:removePeriod} then implies that $\sub$ covers $f$).
    It follows from the above discussion that the loop will be initialized successfully.
    
    We claim that for every $b' \in [q ..b^*-1]$ the pair $(X'_{b'},Y'_0)$ satisfies all conditions checked by the loop.
    \begin{enumerate}
        \item $Y'_0 = Y' \neq \EndSymbol$.
        \item $X'_{b'} \neq \EndSymbol$ since $X'_{b^*} \neq \EndSymbol$, and $\SufffExtend(\EndSymbol)$ is undefined. 
        \item $X'_{b'}$ is a border since $X'_{b^*}=X$ is a border and $X'_{b'}$ can be obtained from $X$ by removing a suffix of length $\rho = \per(X)$ a certain number of times (while remaining with a string with length at least $\rho$).
        Due to \cref{lem:removePeriod}, in every such step we are left with a string covering the previous one.
        In particular, a cover of a border must be a border.
        \item $|X'_{b}| + |Y'_{0}| \le m$. Since for every $b'<b^*$ such that $X'_{b^*} \neq \EndSymbol$ and $X'_{b'} \neq \EndSymbol$ we have $|X'_{b'}| < |X'_{b^*}|$, and we have $|X| + |Y| = |X'_{b^*}| + |Y'_{s^*}| \le m$.
        \item $(X'_0,Y'_{s'})$ is a $2$-cover due to \cref{obs:extendbordersubstring}.
    \end{enumerate}
   It follows that the iterator $b$ of the loop will reach the value $b^*$.
   A similar analysis shows that when reaching $b^*$ and starting the process of increasing $s$, the $s$ iterator of the loop would reach the value $s^*$.
   At this point, the pair $(X,Y) = (X'_{p^*}, Y'_{s^*})$ is reported, as required.

\para{Complexity.}
We analyze the complexity of the nested loop executed in the case in which both $X$ and $Y$ are short periodic. 
The complexity of the rest of the cases arises from the same arguments.
When processing a pair $(X,Y)$, each step of every loop is dominated by the query to the oracle - Obtaining $\PrefExtend^b(X)$ or $\SubExtend^s(Y)$ can be done in constant time using $\PrefExtend^{b-1}(X)$ or $\SubExtend^{s-1}(Y)$ and the rest of the checks consist of integer comparisons.
There is also an additional cost for initializing the loop - extracting $q$ and $f$ from $\Dict_q$ and from $\Dict$, and applying internal pattern matching.
Each of these operations is carried out in constant time.
Each $2$-oracle query on a pair $(X_b,Y_s)$ to the oracle that reports True can be charged on the reported pair $(X_b,Y_s)$.
Due to the uniqueness of the core pair $(X,Y)$ with respect to $(X_b,Y_s)$, the total number of such queries is $\Output$.
Each query on a pair $(X_b,Y_s)$ that reports False is charged on the pair $(X_b,Y_0)$.
If $s > 0$, it means that the pair $(X_b,Y_0)$ was reported as a $2$-cover.
It follows that the total number of such queries on pairs $(X_b,Y_s)$ with $s>0$ is bounded by $\Output$ across all core pairs.
If $s=0$, the loop terminates after the query.
It follows that there is at most one such query for every processed core pair, and their total number across all pairs is bounded by $\Output$ as well.
It follows that the overall complexity of processing all pairs is bounded by $O(\Output \cdot \log^3)$ as required.

In conclusion, the algorithm reports all prefix-suffix $2$-covers and border substring $2$-covers with the required length bound. 
The overall running time, including the preprocessing step dominated by constructing the $2$-cover oracle, is $O(n \log^5 n + \Output \cdot \log^3 n)$, as required.
\end{proof}

\end{document}